\documentclass{aa}  
\usepackage{amsmath}
\usepackage{orcidlink}
\usepackage{natbib}
\usepackage{float}
\usepackage{graphicx}
\usepackage{mwe}
\usepackage{txfonts}
\usepackage{hyperref}
\hypersetup{
     colorlinks   = true,
     citecolor    = blue
}

\usepackage{placeins}
\usepackage{siunitx}
\usepackage{gensymb}
\usepackage{caption,subcaption}
\usepackage{subcaption}
\usepackage{pgffor}
\usepackage{dblfloatfix}    % To enable figures at the bottom of page
\usepackage{lscape} % for 'landscape' environment
\usepackage{multirow}
\usepackage[export]{adjustbox}
\usepackage{numprint}
\npthousandsep{\,}

\usepackage[normalem]{ulem}

\usepackage{tabularx}

\newcommand\clearrow{\global\let\rowmac\relax}
\clearrow

\def \magperarcsec{mag arcsec$^{-2}$}
\def \Hi{\ion{H}{i}}

\def \sigmasfr{$\Sigma_{\rm SFR}$}
\def \msunkpcsq{M$_{\odot}$ kpc$^{-2}$}
\def \msunyrkpcsq{M$_{\odot}$ yr$^{-1}$ kpc$^{-2}$}
\def \halpha{H$\alpha$}
\def \hbeta{H$\beta$}
\def \oiii{[\ion{O}{iii}]$_{5007}$}
\def \nii{[\ion{N}{ii}]$_{6583}$}

\newcommand{\siiab}{[\ion{S}{ii}]$\lambda\lambda$6716,6731}
\newcommand{\niiab}{[\ion{N}{ii}]$\lambda\lambda$6548,6583}

\begin{document}

\title{MUSE observations of the giant low surface brightness galaxy Malin 1: Numerous HII regions, star formation rate, metallicity, and dust attenuation}

    \titlerunning{MUSE observations of the giant low surface brightness galaxy Malin 1}

   \author{Junais\inst{1},
            P.\ M.\ Weilbacher\inst{2},
            B. Epinat\inst{3,4},
            S. Boissier\inst{3},
            G. Galaz\inst{5},
            E. J. Johnston\inst{6},
            T.~H. Puzia\inst{5},
            P. Amram\inst{2},
            K. Ma\l{}ek\inst{1,2}
            }
    \authorrunning{Junais et al.}
  \institute{National Centre for Nuclear Research, Pasteura 7, PL-02-093 Warsaw, Poland\\ %1
             \email{junais@ncbj.gov.pl}
             \and
             Leibniz-Institut f\"ur Astrophysik Potsdam (AIP), An der Sternwarte 16, 14482 Potsdam %2
             \and
             Aix Marseille Univ, CNRS, CNES, LAM, Marseille, France   %3
             \and
             Canada-France-Hawaii Telescope, 65-1238 Mamalahoa Highway, Kamuela, HI 96743, USA %4
             \and
             Instituto de Astrof\'isica, Pontificia Universidad Católica de Chile, Avenida Vicuña Mackenna 4860, 7820436 Macul, Santiago, Chile %5
             \and 
             Instituto de Estudios Astrofísicos, Facultad de Ingeniería y Ciencias, Universidad Diego Portales, Av. Ejército Libertador 441, Santiago, Chile %6
             }

   \date{Received 07 August 2023/ Accepted 18 October 2023}

% \abstract{}{}{}{}{} 
% 5 {} token are mandatory
 
  \abstract
  % context heading (optional)
%   {} leave it empty if necessary  
  {Giant low-surface brightness (GLSB) galaxies are an extreme class of objects with very faint and extended gas-rich disks. Malin 1 is the largest GLSB galaxy known to date, and one of the largest individual spiral galaxies observed so far, but the properties and formation mechanisms of its giant disk are still poorly understood.}
  {We use VLT/MUSE IFU spectroscopic observations of Malin 1 to measure the star formation rate, dust attenuation, and gas metallicity within this intriguing galaxy.
  }
  {We performed a pPXF modeling to extract emission line fluxes such as \halpha{}, \hbeta{}, \nii{} and \oiii{} along the central region as well as the extended disk of Malin 1.}
 {Our observations reveal, for the first time, the presence of strong \halpha{} emission distributed across numerous regions along the extended disk of Malin 1, reaching up to radial distances of $\sim$100 kpc, indicating recent star formation activity. We made an estimate of the dust attenuation in the disk of Malin 1 using the Balmer decrement and found that Malin 1 has a mean \halpha{} attenuation of 0.36 mag. We observe a steep decline in the radial distribution of star formation rate surface density (\sigmasfr{}) within the inner 20 kpc, followed by a shallow decline in the extended disk. 
 We estimated the gas phase metallicity in Malin 1, and also found for the first time, that the metallicity shows a steep gradient in the inner 20 kpc of the galaxy from solar metallicity to sub-solar values, followed by a flattening of the metallicity in the extended disk with a relatively high value of $\sim$0.6 $Z_{\odot}$. We found that the normalized abundance gradient of the inner disk of Malin 1 is similar to values found in normal galaxies. However, the normalized gradient observed in the outer disk can be considered extreme when compared to other disk galaxies.
 A comparison of the star formation rate surface density and gas surface density shows that, unlike normal disk galaxies or other LSBs, the outer disk of Malin 1 exhibits relatively low star formation efficiency based on atomic gas mass estimates, which may be mildly exacerbated by the vanishing upper molecular gas mass limits found by recent CO studies.
 }
 {Owing to the detection of emission lines over a large part of the Malin 1 extended disk, this work sheds light on the star formation processes in this unique galaxy, highlighting its extended star-forming disk, dust attenuation, almost flat metallicity distribution in the outer disk, and exceptionally low star-formation efficiency. Together with previous results, our findings contribute to a more detailed understanding of the formation of the giant disk of Malin 1 and also to constrain possible proposed scenarios on the nature of GLSB galaxies in general.
 }

   \maketitle
%
%-------------------------------------------------------------------

\section{Introduction}

Low surface brightness galaxies (LSBs) form a diverse class of galaxies that exhibit significantly lower brightness per unit area than ``normal'' high surface brightness galaxies, and certainly lower surface brightness than the dark night sky. In the past decade, LSBs have obtained a lot of attention due to their extreme characteristics and potential implications for our understanding of galaxy formation scenarios. LSBs are commonly defined as galaxies with an average \textit{r}-band surface brightness ($\bar{\mu}_{r}$) below the typical level of the night sky ($\bar{\mu}_{r} > 23$ \magperarcsec{}; \citealt{martin2019,Junais2023}). Among LSBs, there is a distinct sub-population known as giant LSB galaxies (GLSBs), having a massive faint extended disk and rich in gas content \citep{Bothun1987,Sprayberry1995,Matthews2001}.

Malin 1 is the archetype of GLSB galaxies and has captivated the attention of astronomers since its accidental discovery nearly four decades ago \citep{Bothun1987}. Malin 1 has a radial extent of at least $\sim$120 kpc \citep{moore2006}, and an extrapolated central disk central surface brightness of $\mu_{0,V}\approx25.5 {,,}\text{mag arcsec}^{-2} $ \citep{Impey1997}. Despite its faint surface brightness disc, Malin 1 has a total absolute magnitude of $M_V \approx -22.9$ mag \citep{pickering1997} and an HI mass of approximately $5 \times 10^{10} M_{\odot}$ \citep{pickering1997,Matthews2001}. It is currently considered the largest spiral galaxy known to date. Malin 1 is situated in a relatively low-density environment in the large-scale structure with close proximity to a filament, offering the stability and richness of its huge gaseous disk \citep{junais2021_thesis}.

A Hubble Space Telescope (HST) \textit{I}-band image analysis by \citet{barth2007} has identified that Malin 1 has a normal barred inner spiral disk embedded within an extensive, diffuse LSB disk. This has similarities to galaxies with extended ultraviolet (XUV) discs found in approximately 30\% of nearby galaxies \citep{thilker2007}. The extension and scope of the enormous spiral arms of Malin 1 were shown by \citet{galaz2015}. Later \citet{boissier2016} performed an analysis on the radial stellar profiles and suggested that the extended disk of Malin 1 has an angular momentum about 20 times larger than Milky Way. Therefore, Malin 1 represents an extreme case among the class of GLSBs. The nature and origin of such GLSBs are still poorly understood, although more GLSBs, but less extreme ones, were discovered over the years \citep{hagen2016,Saburova2021}. In recent work, \citet{Saburova2023} suggested that GLSBs are not rare objects as previously thought. Based on the volume density of GLSBs they predicted, $\sim$13000 GLSBs could exist within the local universe at $z<0.1$. It is expected that soon, with the deep Legacy Survey of Space and Time \citep[LSST; ][]{ivezic2019}, we will be able to uncover even more of these giant, faint galaxies.

Despite its significance, Malin 1 has been subject to limited spectroscopic studies. Previous efforts have primarily relied on velocity maps with low spatial resolution obtained from HI data \citep{Lelli2010}, along with optical spectra of the central region from SDSS observations \citep{Subramanian2016}. Later, \citet{Junais2020} performed a spectroscopic analysis of Malin~1 using long-slit data obtained from the IMACS/Magellan spectrograph, focusing mostly on the central region (<10 kpc) of Malin~1, along with only one small region detected in the extended disk (at $\sim$25 kpc radius). The lack of in-depth spectroscopic analysis of Malin 1, especially on its large extended disk, hinders a comprehensive understanding of its properties and formation. 

In this work, we present a spectroscopic study of Malin 1 utilizing MUSE Integral Field Unit (IFU) data, aiming to shed light on the nature and formation of this extraordinary GLSB galaxy. Based on these data we perform a detailed analysis of Malin~1's star formation rate, dust attenuation, and gas-phase metallicity of its extended disk.
Throughout this work, we adopt a flat $\Lambda$CDM cosmology  with $H_0 = 70 \,\,\text{km s}^{-1} \text{Mpc}^{-1}$,  $\Omega_{M} = 0.27$ and $\Omega_{\Lambda} = 0.73$, which corresponds to a projected angular scale of $1.56 \,\,\text{kpc arcsec}^{-1}$ and a luminosity distance of 377 Mpc. 

The paper is structured as follows: Sect. \ref{sect:data} provides an overview of the data and observations, including the MUSE IFU data acquisition and reduction process. Section \ref{sect:results} focuses on the analysis and results. Section \ref{sect:discussion} and \ref{sect:conclusions} present the discussion and the conclusions, respectively.

\section{Observation and Data Analysis}\label{sect:data}

\begin{figure}
    \centering
    \includegraphics[width=0.49\textwidth]{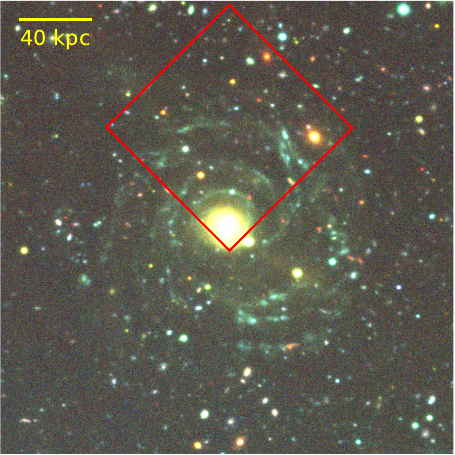}
    \caption{Colour composite image of Malin 1 from the CFHT-Megacam NGVS (\citealt{ferrarese2012}) \textit{u}, \textit{g,} and \textit{i}-band images. The image spans a width of $\sim2.6\arcmin \times 2.6\arcmin$. The red box shows the MUSE field of observation ($1\arcmin \times 1\arcmin$). 
    }
    \label{fig:cfht_image}
\end{figure}

Malin 1 was observed with the VLT/MUSE integral field spectrograph \citep{Bacon2010} on 18 April 2021 under program ID 105.20GH.001 (PI Gaspar Galaz). The galaxy covers more that $2\arcmin \times 2 \arcmin$ on the sky, but the field of view of MUSE is $1\arcmin \times 1 \arcmin$, with a sampling scale of 0.2\arcsec\ pixel$^{-1}$. Out of the four planned MUSE pointings, only the northern quadrant of Malin 1 was observed (see Fig. \ref{fig:cfht_image}). However, the center of the galaxy was well covered. Observations were conducted at an airmass of $\sim$1.3 and an external 
seeing of $\sim$1.2\arcsec. The observations were carried out with four exposures resulting in a total exposure time of 4640 seconds. 
The spectral resolving power ranges from $R\!\simeq\!1770$ at 4800 $\AA$ to $R\!\simeq\!3590$ at 9300 $\AA$.
The observations used the ground-layer adaptive optics (AO) system \citep{Stroebele2012} so that the full width at half maximum (FWHM) in the MUSE instrument at 7000\,\AA{} was estimated to be $\sim$0\farcs55 from the telemetry of the AO system \citep{Fusco2020}. % 

\subsection{Data reduction}

We reduced the data with the MUSE pipeline (v2.8, \citealt{Weilbacher2020}) called from the ESO Recipe Execution Tool (EsoRex) tool. We largely used standard processing, including the creation of master bias, flat field, and trace tables, as well as deriving wavelength solutions separately for each CCD, with an overall twilight-sky correction for the whole field, all with default parameters. We also used standard bad pixel and geometry tables as well as a line-spread function automatically associated with the data by the ESO archive. All master calibrations were then applied to the raw on-sky data (standard stars, science exposures, and offset sky fields). While the standard star (HD\,49798) observed at the beginning of the night was usable for the flux calibration, we used two other standard star exposures to derive the telluric correction (LTT\,3218 from 14 April 2021 and EG\,274 taken on 14 April 2021). These provide a good match in the integrated water vapor levels (recorded in the IWV keywords in the raw data headers) and hence correct the telluric A- and B-bands better. The science data was corrected for the sky background, using internal sky instead of the offset sky fields, since the spiral arms of Malin 1 fill only a portion of the MUSE field. 
The data were further corrected for distortions using the provided astrometric calibration and shifted to a barycentric velocity frame. All four science exposures were aligned using a star from the GaiaEDR3 catalog \citep{Lindegren2021} and combined to form a single data cube with wavelength coverage 4595--9350\,\AA. The final spatial FWHM measured in the reconstructed $R$-band is 0\farcs63. % 

\subsection{Emission line fitting}\label{sect:ppxf_fit}

To disentangle the ionized gas emission lines from the stellar continuum and measure their fluxes corrected for hydrogen absorption, throughout this work, we used the Penalized PiXel-Fitting (pPXF) tool \citep[v8.2.4,][]{Cappellari2017}. To maximize the detection along different regions of Malin 1, we employed several approaches for the emission line fitting. This is described in the following sub-sections.

\subsubsection{Central region}\label{sect:ppxf_fit_central_region}

\begin{figure*}
    \centering
    \includegraphics[width=0.49\textwidth]{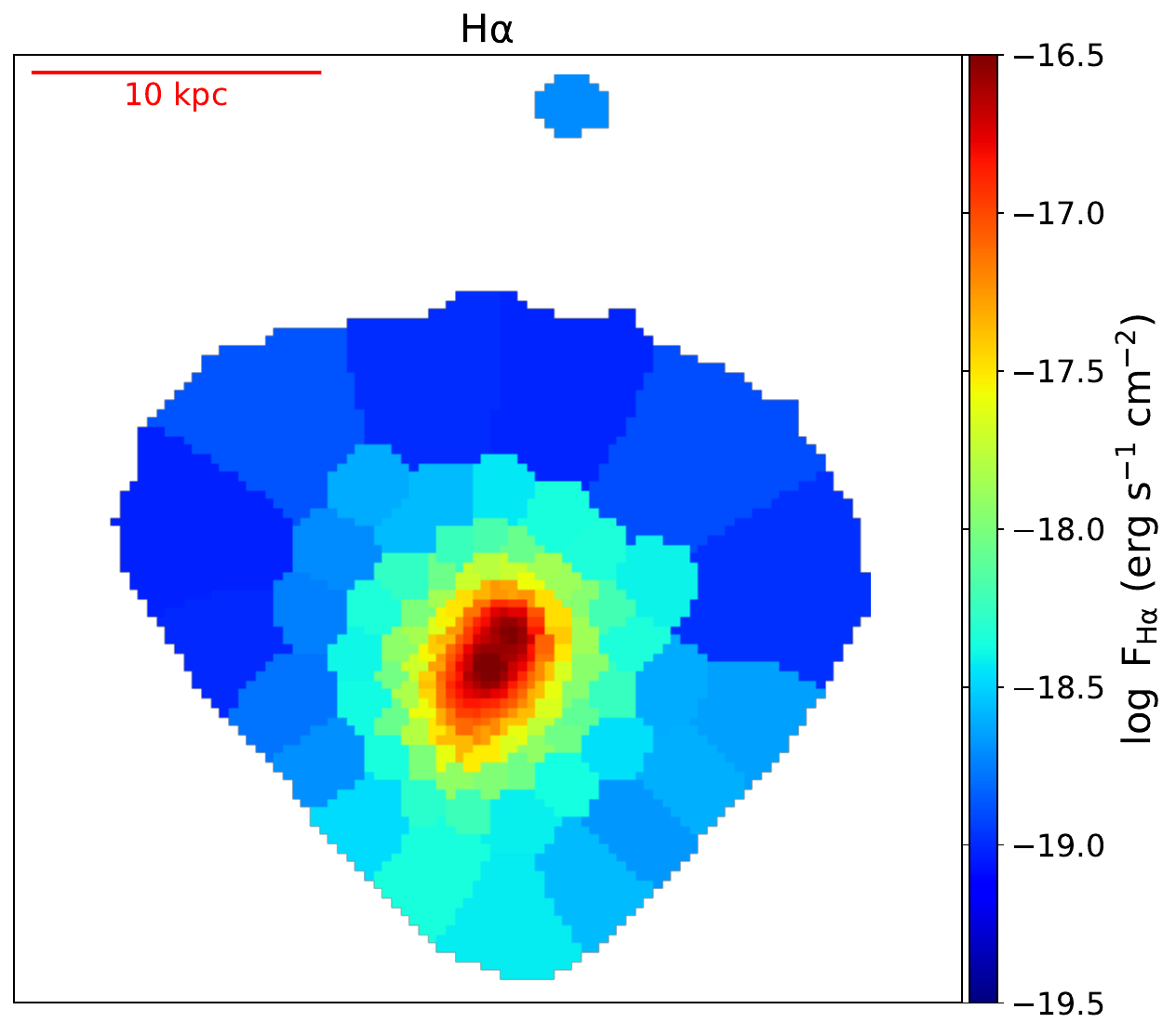}
    \includegraphics[width=0.49\textwidth]{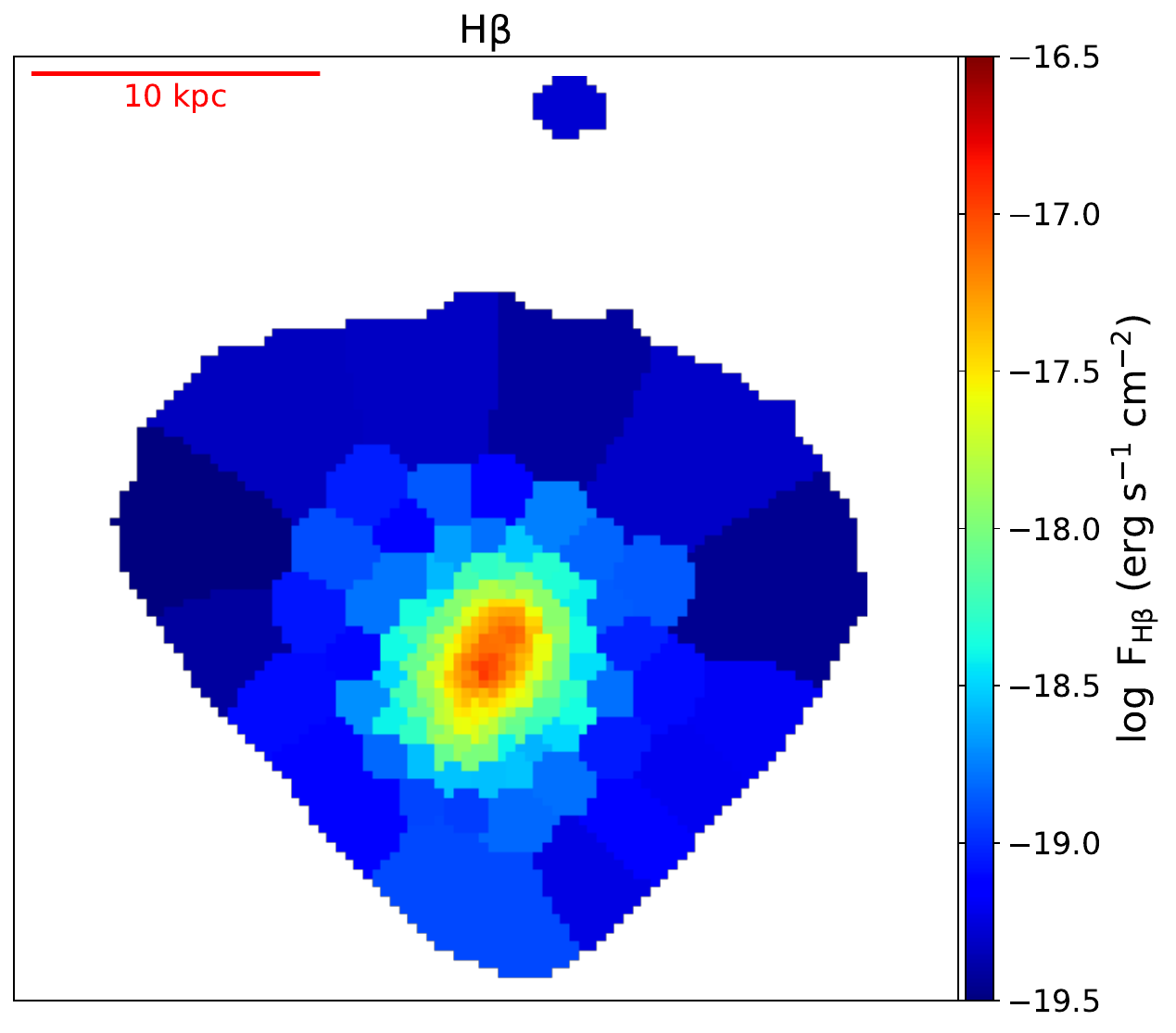}
    \includegraphics[width=0.49\textwidth]{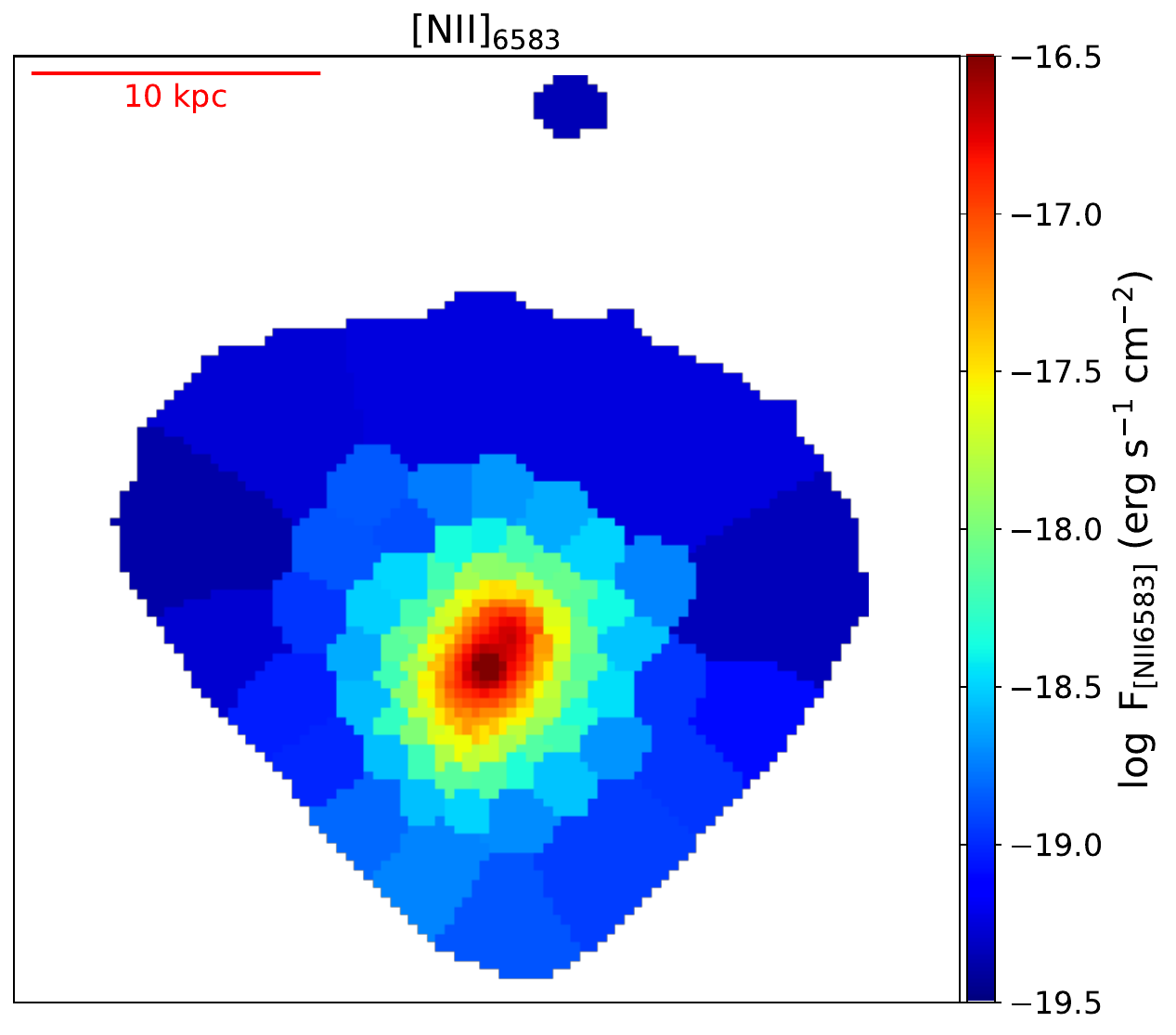}
    \includegraphics[width=0.49\textwidth]{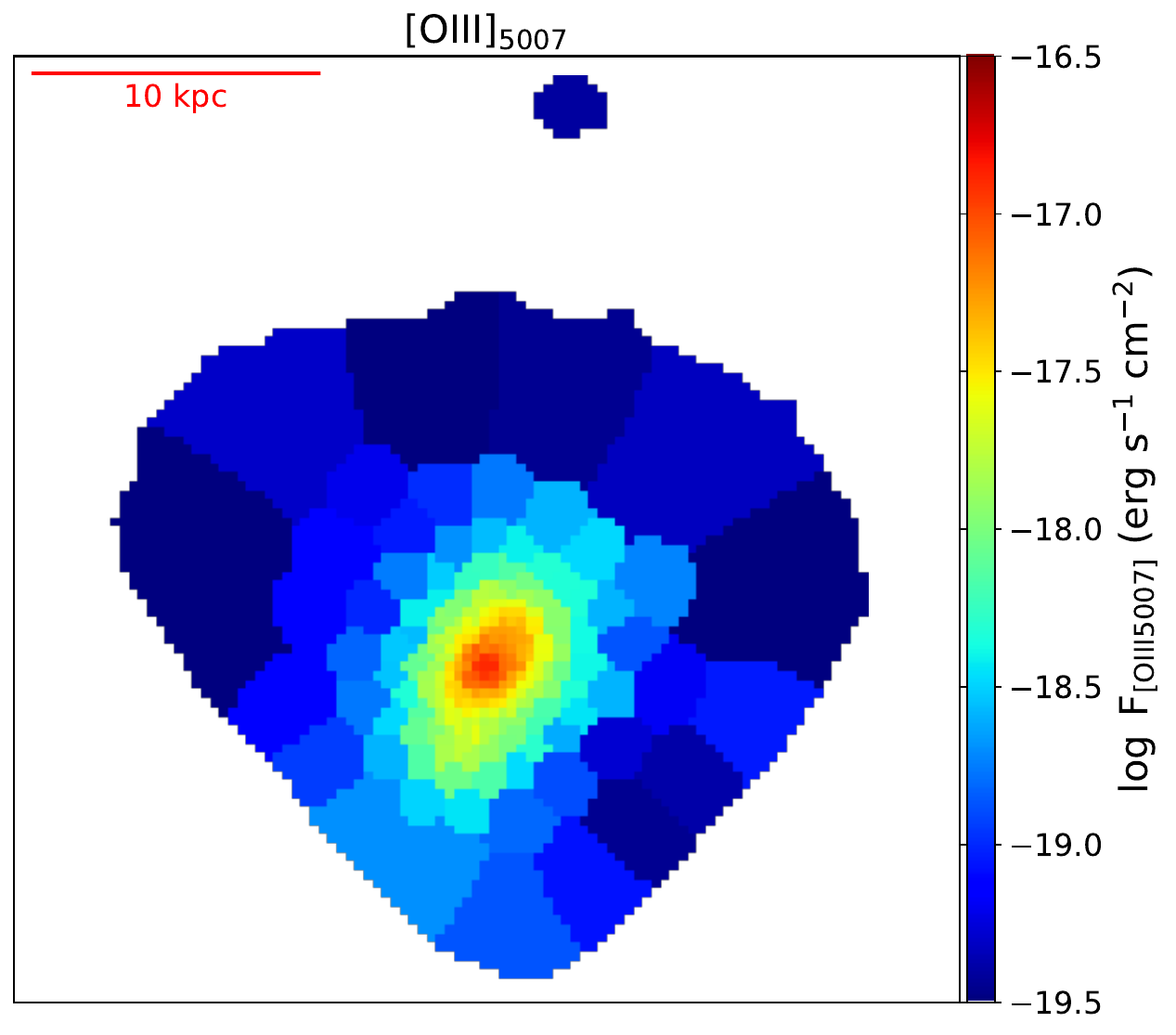}
    \caption{Emission line flux maps of the central region of Malin 1 ($21\arcsec \times 21\arcsec$) obtained from the pPXF fitting. The \halpha{}, and the \hbeta{} lines are in the top panels, whereas the \nii{} and the \oiii{} lines are along the bottom panels. The color bar indicates the flux corresponding to each line.}
    \label{fig:ppxf_maps}
\end{figure*}

In the central region of Malin 1 ($21\arcsec \times 21\arcsec$) where the stellar continuum is strong, we performed a spectral binning using the Voronoi algorithm of \citet{Cappellari2003} to target a final signal-to-noise ratio (SNR) $\approx20$. We include only the spatial pixels (spaxels) where the median continuum SNR in the wavelength range 5600--6500\,\AA{} was at least 1. Then, on the binned spectra of the central region, the pPXF was set up to simultaneously fit the continuum, using the GALEV SSPs \citep{Kotulla2009} constructed with the Munari stellar library \citep{Munari2005} as templates, and the known strong emission lines between H$\gamma$ and Pa14 -- corresponding to the restframe $\lambda$-range at the redshift of Malin 1 -- modeled as Gaussian peaks. To reduce the effects of flux calibration inaccuracies we employ a multiplicative polynomial of order 7. 
This is the same setup already employed and discussed in more detail by \citet{Weilbacher2018} and \citet{Micheva2022}. Lines falling into masked spectral regions\footnote{We mask significant telluric residuals that might affect the fit, and also the NaD-gap created to suppress the AO laser emission, and also the atmospheric Raman features created by the lasers.} at the redshift of Malin 1 were excluded. As initial guesses for the kinematics, we used $z = 0.08$ for the velocity and $\sigma=75$\, km\,s$^{-1}$, for both stars and gas. %
We check this setup against the default setup of pPXF using the E-MILES SSP templates \citep{Vazdekis2016}. Since we do not find significant differences, we prefer to use the results from the GALEV+Munari setup in the following part of this work, as it produces fits that have fewer residuals or artifacts.
All emission line flux measurements are then projected back into the 2D plane to form a map (see Fig.~\ref{fig:ppxf_maps}).

\subsubsection{Extended disk regions}\label{sect:ppxf_fit_extended_disk}

\begin{figure*}
    \centering
    \includegraphics[width=0.49\textwidth]{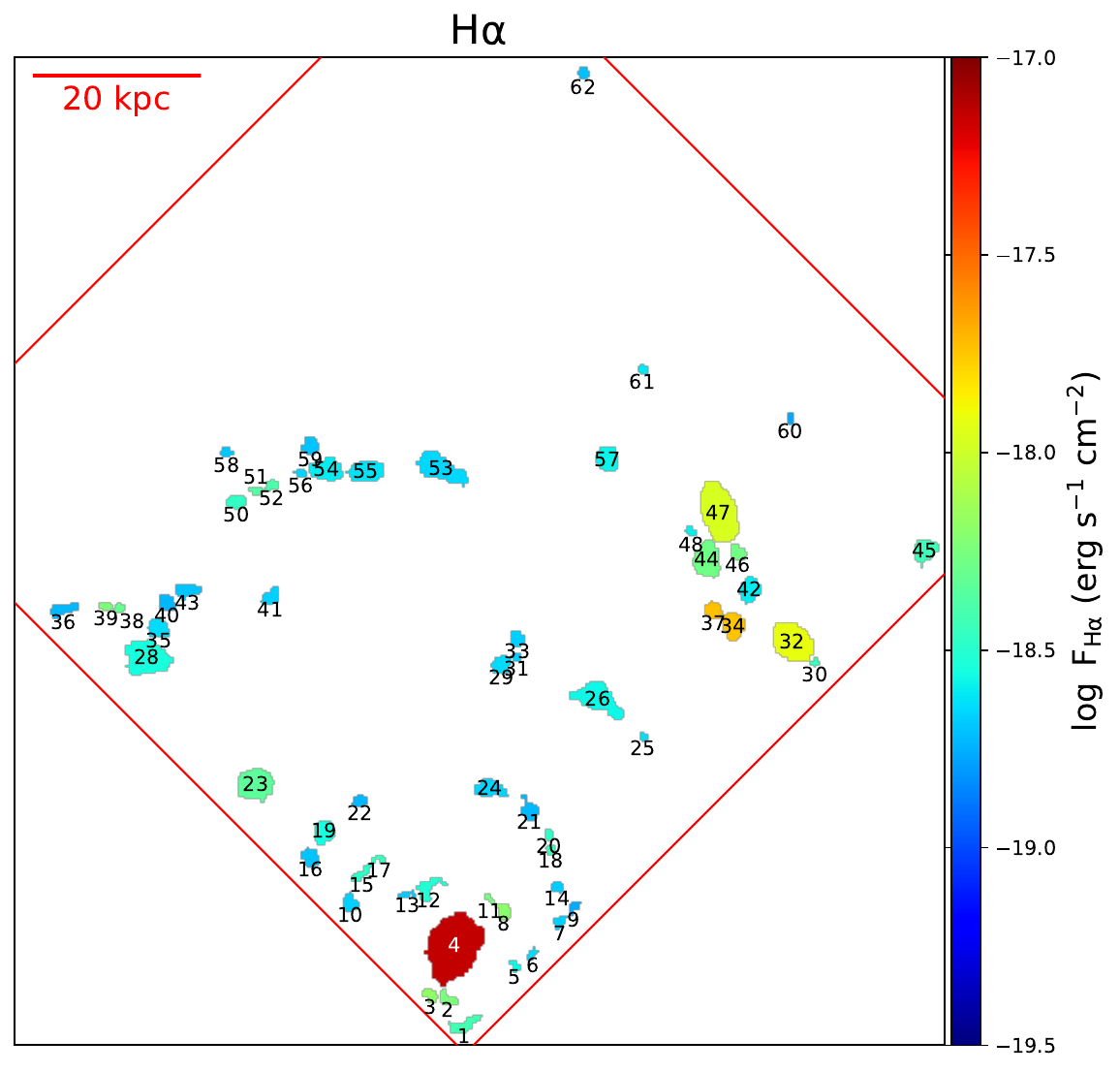}
    \includegraphics[width=0.49\textwidth]{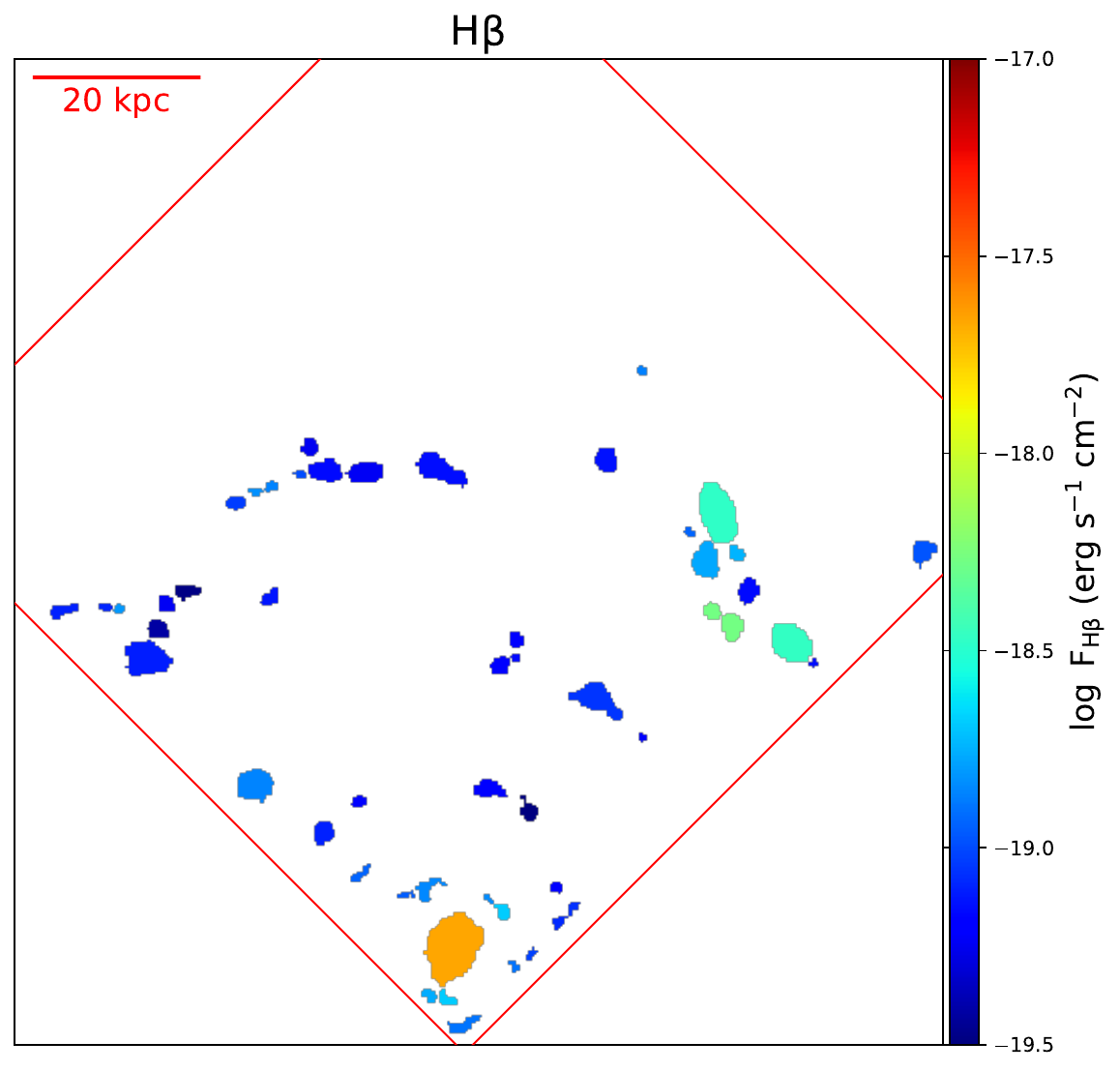}
    \includegraphics[width=0.49\textwidth]{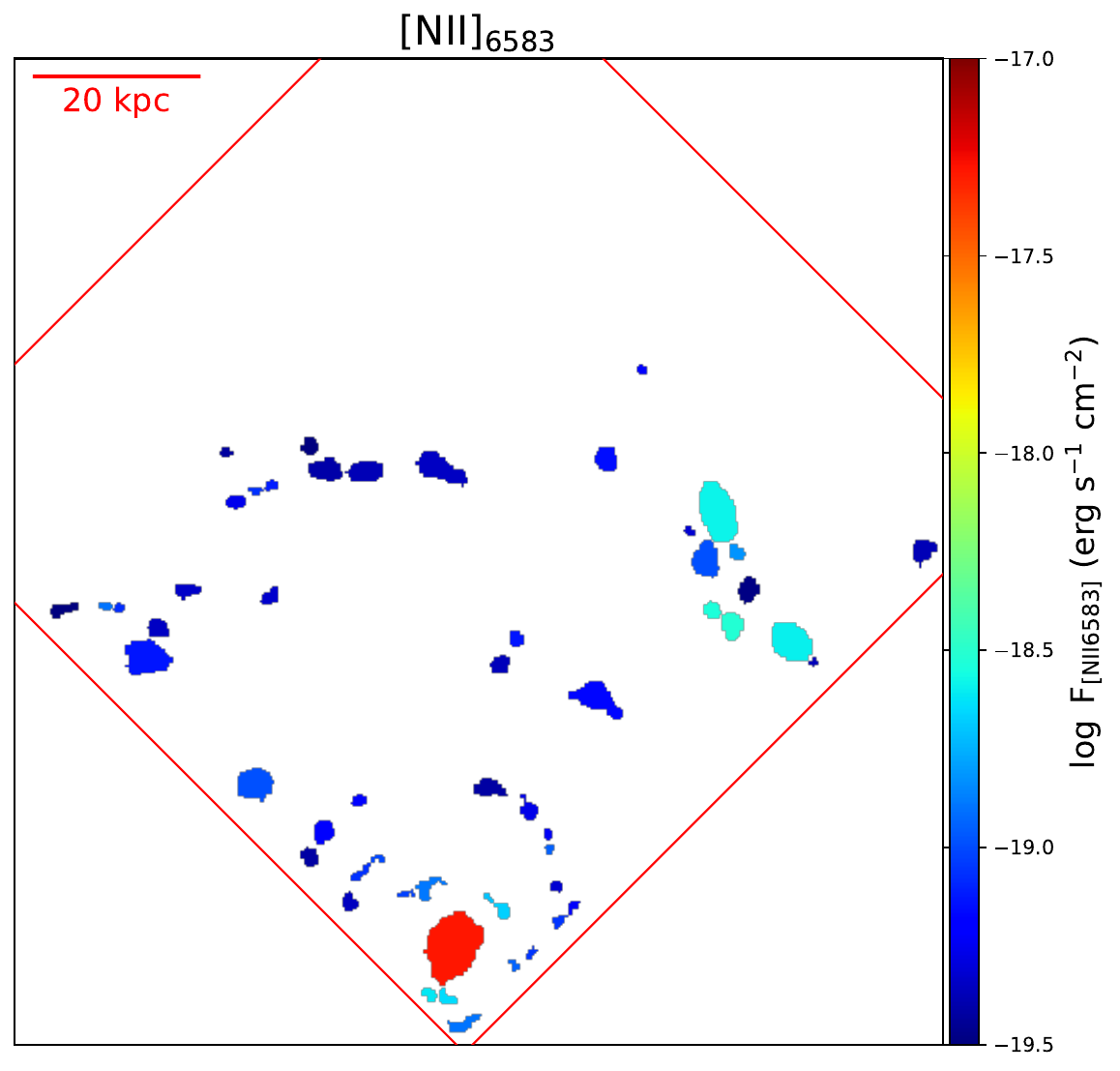}
    \includegraphics[width=0.49\textwidth]{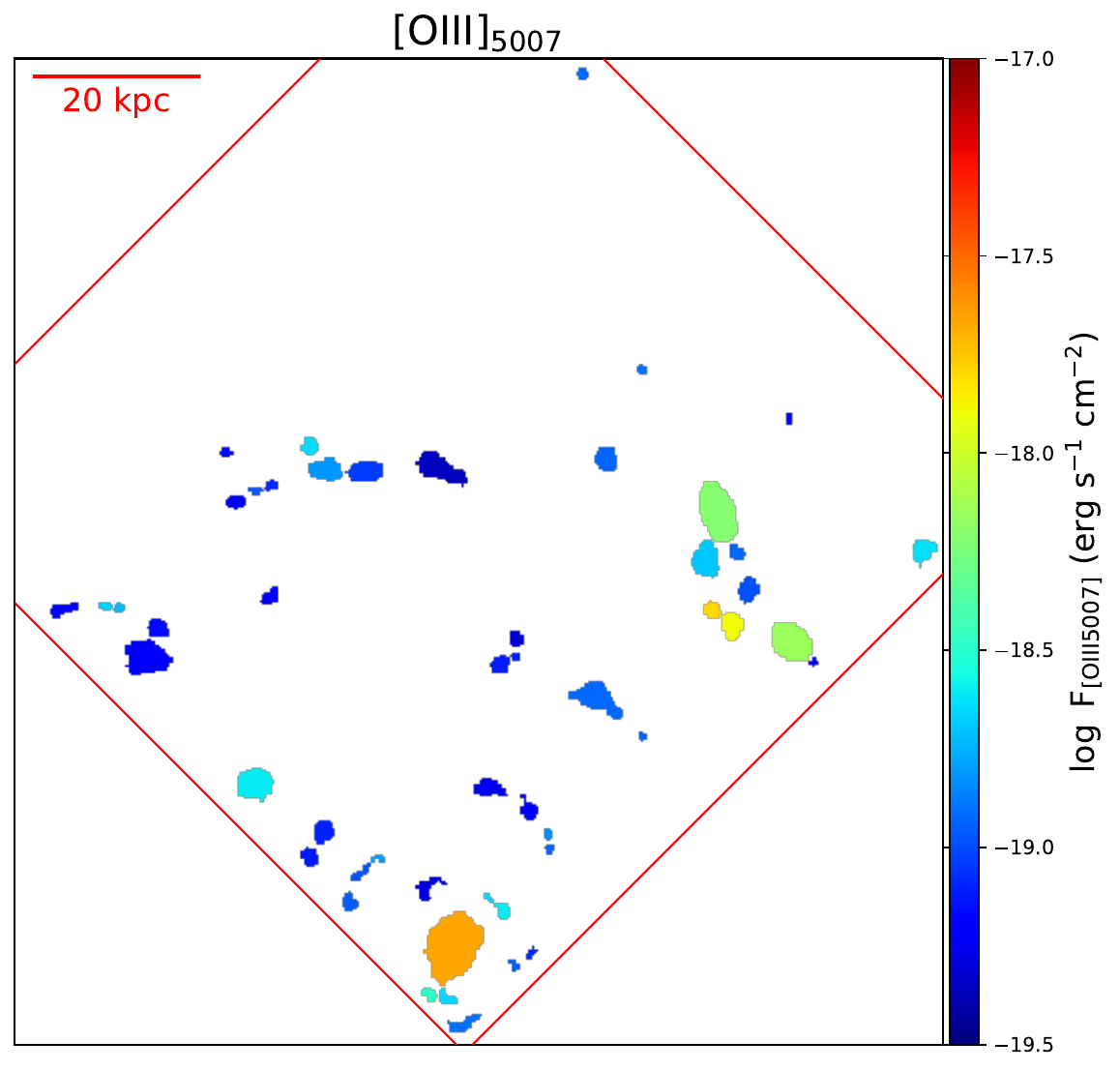}
    \caption{Emission line flux maps of Malin 1 obtained from the pPXF fitting of the HII regions. The \halpha{}, and the \hbeta{} lines are in the top panels, whereas the \nii{} and the \oiii{} lines are along the bottom panels. The ID of each region, as discussed in Sect. \ref{sect:ppxf_fit_extended_disk}, is labeled in black in the top left panel. The regions with ID 27 and 49 are excluded from the maps as they do not have an $\rm SNR > 2.5$ in any of the emission lines. The color bar indicates the flux corresponding to each emission line. Note that the flux value of each region is given as the average flux per pixel within that region obtained from the fitting of its row-stacked spectrum.}
    \label{fig:Gaussian_maps}
\end{figure*}

To extract emission line fluxes over the whole MUSE field, including the extended disk of Malin 1, we first defined HII regions using the dendrogram algorithm (using the Python package \texttt{astrodendro}\footnote{\url{http://www.dendrograms.org/}}) which tracks isophotal contours around peaks in images. It creates a tree-like structure that allows us to relate \textit{leaves} (contours within which only a single peak is located), to \textit{branches} (contours containing several leaves). We ignore the hierarchical nature of the data structure and just use the leaves, which represent the largest contour around a peak that has not merged with neighboring peaks, as the HII regions. As input, we use an H$\alpha$ image created by fitting a Gaussian function to the expected redshifted emission line in each spaxel of the cube. We filter this image with a 2D Gaussian function with 0\farcs6 FWHM to enhance real features. Peaks are detected above a limit of $1.5\times10^{-19}$ erg s$^{-1}$ cm$^{-2}$, with a minimum number of 9 pixels above the limit. We do not impose a limit on the height of the peaks. We detect 62 peaks in the cube and extract spectra and data variance by averaging them over the spaxels of the corresponding leaves (HII regions) defined by \texttt{astrodendro}. They are saved as row-stacked spectra and subsequently input to pPXF as discussed in Sect. \ref{sect:ppxf_fit_central_region} to extract emission line fluxes. Only the line fluxes with an $\rm SNR>2.5$ were used in this analysis. We found that such an approach of fitting the integrated spectrum of a region, compared to a pixel-level fitting, significantly increases the SNR and hence increases the number of regions (by a factor of about 8) with faint emission line measurements (\hbeta{}, \nii{} and \oiii{}). Figure \ref{fig:Gaussian_maps} shows the emission line maps of all the identified regions. We can clearly see several \halpha{} detected regions throughout the disk of Malin 1, with most of them extending up to $\sim 80$ kpc from the center of the galaxy and one region detected at a radius of about 100 kpc (ID 62 in Fig. \ref{fig:Gaussian_maps}).

\subsubsection{\halpha{} radial average}\label{sect:ppxf_halpha_radial_average}

Apart from the fitting of the central region and the selected HII regions in the extended disk, we also performed an integrated spectral fitting along several radial bins of Malin 1. This was done to estimate the \halpha{} radial average fluxes within the galaxy.
For this purpose, we placed 14 concentric rings on the MUSE cube, starting from the center of Malin 1 to a radial distance of $\sim$100 kpc, each ring with a width of 5\arcsec{} (7.8 kpc). Then we followed a similar approach as discussed in Sect. \ref{sect:ppxf_fit_extended_disk}, by stacking all the spaxels within a ring to obtain its corresponding integrated spectrum. However, as each radial bin spans over several kpcs, resulting in azimuthal variations of the line-of-sight velocity, we need to correct for such velocity variations along the spaxels of a ring to generate high SNR integrated spectra\footnote{We did not perform any velocity corrections during the HII region spectral stacking discussed in Sect. \ref{sect:ppxf_fit_extended_disk}, as the average size of a region ($\sim$2 kpc) in that case was small to have any significant velocity variations.}. 

To correct for the velocity variations, we extracted moment maps and masks over the whole MUSE field-of-view using the python software \textsc{Camel}\footnote{\url{https://gitlab.lam.fr/bepinat/CAMEL}} \citep[see][]{Epinat2012} on the group of emission lines \halpha{}, \siiab{}, and \niiab{}. In order to avoid local velocity variations and to increase the sensitivity at the edge of emitting regions, and thus the spatial extent, the MUSE datacube was first smoothed using a 2-pixel FWHM Gaussian kernel. Only the spaxels within the wavelength range corresponding to the redshift range $0.0714<z<0.0934$ are considered around each line within the MUSE cube in order to have some continuum but with a reasonable weight in the fit. \textsc{Camel} then fits all lines and the continuum simultaneously for each spaxel of the MUSE cube, taking advantage of the variance cube produced during data reduction.
Lines are modeled as Gaussian functions, using a common velocity but with a width that can vary from one species to another, since both lines in a doublet are expected to have the same origin and are close enough in wavelength to avoid line-spread function variations. The continuum is adjusted with a third-degree polynomial function. Flux maps are generated for all fitted emission lines, together with associated error maps, as well as SNR maps, velocity dispersion fields, and the velocity field.
We further compute a spatial mask in order to exclude regions with no signal coherent with the large-scale velocity pattern. We remove all pixels having an SNR in the \halpha{} line below 2.5 and with velocities incompatible with Malin 1, and further removed spurious isolated groups of pixels smaller than 1\arcsec{} in diameter, which is smaller than the seeing FWHM of the observations.

Once those maps are generated, for each spaxel of the original unsmoothed datacube, the inferred velocity is used to compute the wavelengths at which the spectrum is actually sampled in the restframe corresponding to Malin 1 systemic velocity. The spectrum at each spaxel is then re-sampled on a single spectral grid for all spaxels by performing a linear interpolation to apply the velocity correction. Finally, for each ring, we sum all spaxels at Malin 1 rest, lying within both the ring and mask to produce the corresponding integrated spectrum. We performed a pPXF fit on these spectra as discussed in Sect. \ref{sect:ppxf_fit_central_region} to obtain the radial average \halpha{} fluxes along each ring (the average \halpha{} flux along each ring was obtained by dividing the total flux by the unmasked ring area). We only include the \halpha{} fluxes with an $\rm SNR>2.5$ in this work (11 among the 14 rings). 

From hereupon, we correct all the observed emission line fluxes from Sect. \ref{sect:ppxf_fit} for foreground Galactic extinction using the \citet{Schlegel1998} dust maps and the \citet{Cardelli1989} Milky Way dust extinction law.

\section{Results}\label{sect:results}

\subsection{Balmer decrement and dust attenuation} \label{sect:balmer_ratio}

The Balmer ratio (\halpha{}/\hbeta{} flux ratio) is commonly used as a diagnostic tool for dust attenuation in galaxies \citep[e.g.,][]{dominguez2013,Boselli2015}. The intrinsic Balmer ratio, (\halpha{}/\hbeta{})$_{\rm int}$, remains roughly constant for typical gas conditions in galaxies. For Case B recombination\footnote{This assumes 1) optically thin gas, 2) which is ionized by a harder radiation field than produced by the recombination process itself, and 3) that there is negligible influence of the ionizing radiation on the gas temperature. See also \cite{Nerbin2023}.}, (\halpha{}/\hbeta{})$_{\rm int} = 2.86$ \citep{Osterbrock1989}. Therefore, comparing the observed Balmer ratio with the theoretical value, we can obtain the attenuation at a wavelength, $A_{\lambda}$, following Eq. 6 of \citet{Yuan2018}:

\begin{equation}\label{eq:balmer_attenuation}
    A_{\lambda} = -2.5\frac{k(\lambda)}{k(\mathrm H\alpha) - k(\mathrm H\beta)} \log\Biggr[\frac{(\mathrm H\alpha/\mathrm H\beta)_{\rm obs}}{2.86}\Biggr], 
\end{equation}

where $k(\lambda)$ is the value of the attenuation curve at a wavelength $\lambda$ and $(H\alpha/\mathrm H\beta)_{\rm obs}$ is the observed Balmer ratio. Assuming a \citet{calzetti2000} dust attenuation law with $R_V = 4.05$, we obtain $k(\mathrm H\alpha) = 3.33$; $k(\mathrm H\beta) = 4.60$ ; 
$k(\mathrm [NII]_{6583}) = 3.31$ and $k(\mathrm [OIII]_{5007}) = 4.46$. We can then obtain the attenuation for all the emission lines presented in this work using Eq. \ref{eq:balmer_attenuation}.

\begin{figure*}
    \centering
    \includegraphics[width=0.49\textwidth]{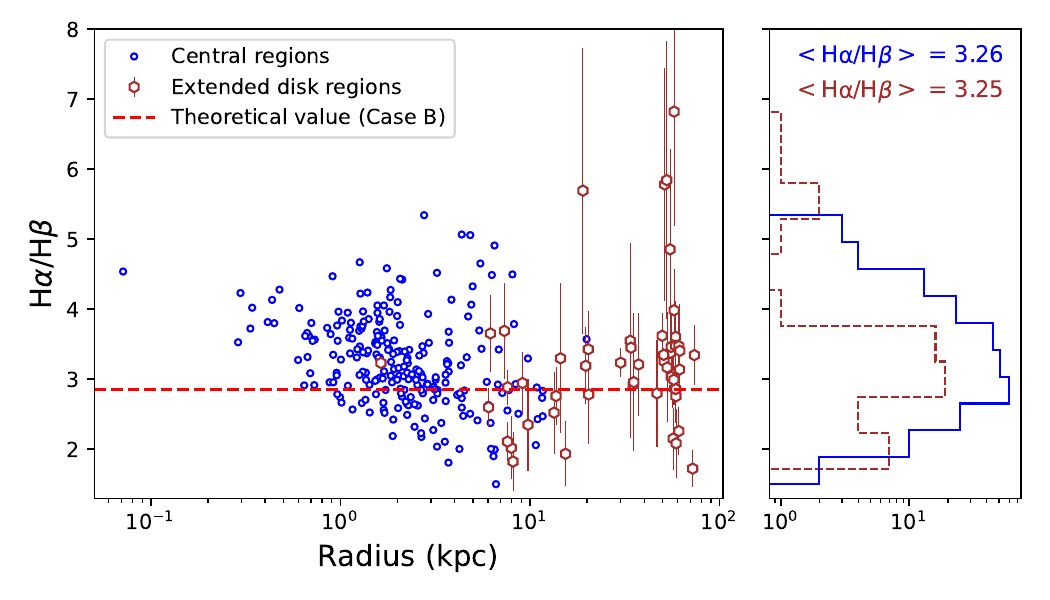}
    \includegraphics[width=0.49\textwidth]{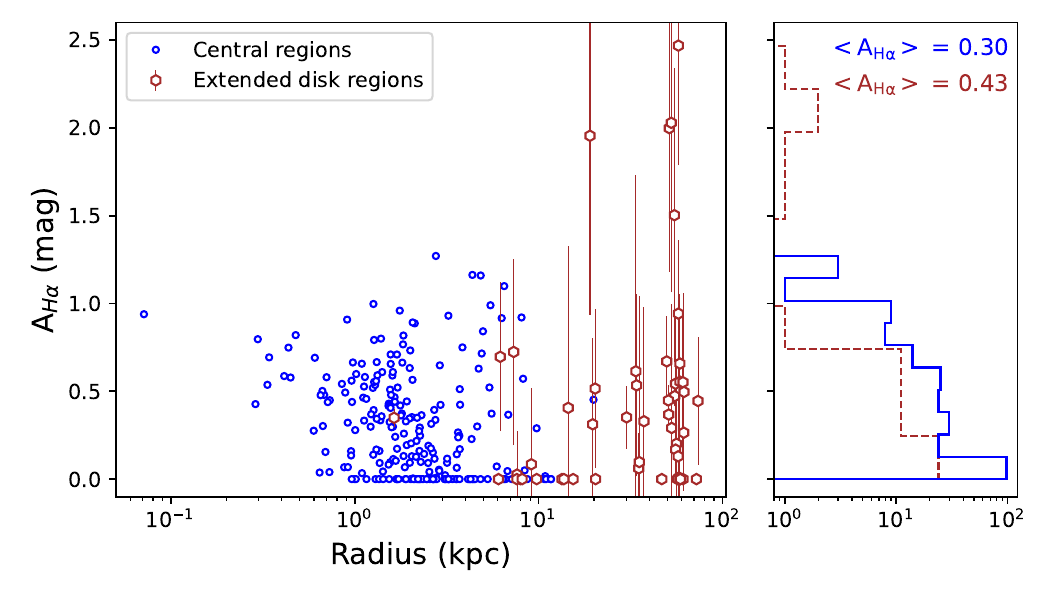}
    \caption{Radial variation of Balmer ratio (left panel) and \halpha{} attenuation (right panel). The blue circles and the brown hexagons are the central regions and the extended disk regions, respectively, obtained from the pPXF fit discussed in Sect. \ref{sect:ppxf_fit}. The red horizontal dashed line marks the intrinsic Balmer ratio of 2.86 for Case B recombination.
    To all regions with the Balmer ratio below this value, we assign zero attenuation. The histograms beside each panel give the overall distribution of each quantity (blue solid line and brown dashed line for the central region and extended disk, respectively), with their mean values indicated at the top of each panel.}
    \label{fig:balmer_ratio_attenuation}
\end{figure*}

Fig. \ref{fig:balmer_ratio_attenuation} shows our observed Balmer ratio and \halpha{} attenuation. We can see that the central region of Malin 1 ($<10$ kpc) has a large range of Balmer ratios up to 5, with a mean value of about 3.26. This corresponds to an \halpha{} attenuation ($A_{H\alpha}$) up to 1 mag, with a mean attenuation of $\sim$0.3 mag. Similar to the central region, in the extended disk we observe a mean Balmer ratio of 3.25 and $A_{H\alpha}$ of 0.43 mag. We thus conclude that Malin 1 has non-negligible dust attenuation in the central and extended disk regions with a mean $A_{H\alpha}=0.36$ mag.

\subsection{Star formation rate surface density}\label{sect:sigma_sfr}

We use our \halpha{} emission line flux measurements to estimate the radial variation of the star formation rate surface density (\sigmasfr{}) in Malin 1. The measured \halpha{} fluxes were converted to star formation rate (SFR) following \citet{Boissier2013} and a \citet{Kroupa2001} initial mass function (IMF) using

\begin{equation}\label{eqn:sfr_halpha}
    \rm SFR_{\rm H\alpha} \,(\rm M_{\odot} \,yr^{-1}) = 5.1\times10^{-42} \,L_{\rm H\alpha} \,(\rm erg \,s^{-1}),
\end{equation}

where $L_{\rm H\alpha}$ is the \halpha{} luminosity corresponding to the observed \halpha{} flux. The \halpha{} fluxes, before the estimation of SFR, were corrected for dust attenuation using the attenuation measurements discussed in Sect. \ref{sect:balmer_ratio}. The SFR values were converted to \sigmasfr{} by dividing by the area in which the \halpha{} flux was measured.

\begin{figure}
    \centering
    \includegraphics[width=0.49\textwidth]{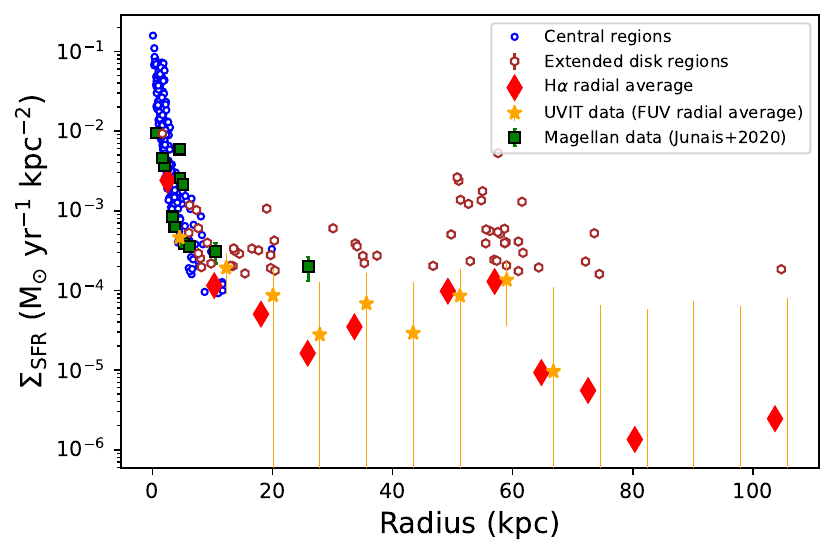}
    \caption{Star formation rate surface density of Malin 1 as a function of galactocentric radius. The blue circles and the brown hexagons are the central regions and the extended disk regions, respectively, obtained from the pPXF fit discussed in Sect. \ref{sect:ppxf_fit}. The red diamonds are the \halpha{} radial average measured along concentric rings of 5\arcsec{} width from the center. The orange stars are the radial averages measured on the same field in the UVIT FUV image of Malin 1 from \citet{Saha2021}, as shown in Fig.~\ref{fig:uvit_fuv}. For illustration purposes, the UV data points are horizontally shifted by 2~kpc. The green squares are the data points from \citet{Junais2020}, based on the IMACS-Magellan \halpha{} long-slit spectra of Malin 1.}
    \label{fig:sigmasfr}
\end{figure}

\begin{figure}
    \centering
    \includegraphics[width=0.49\textwidth]{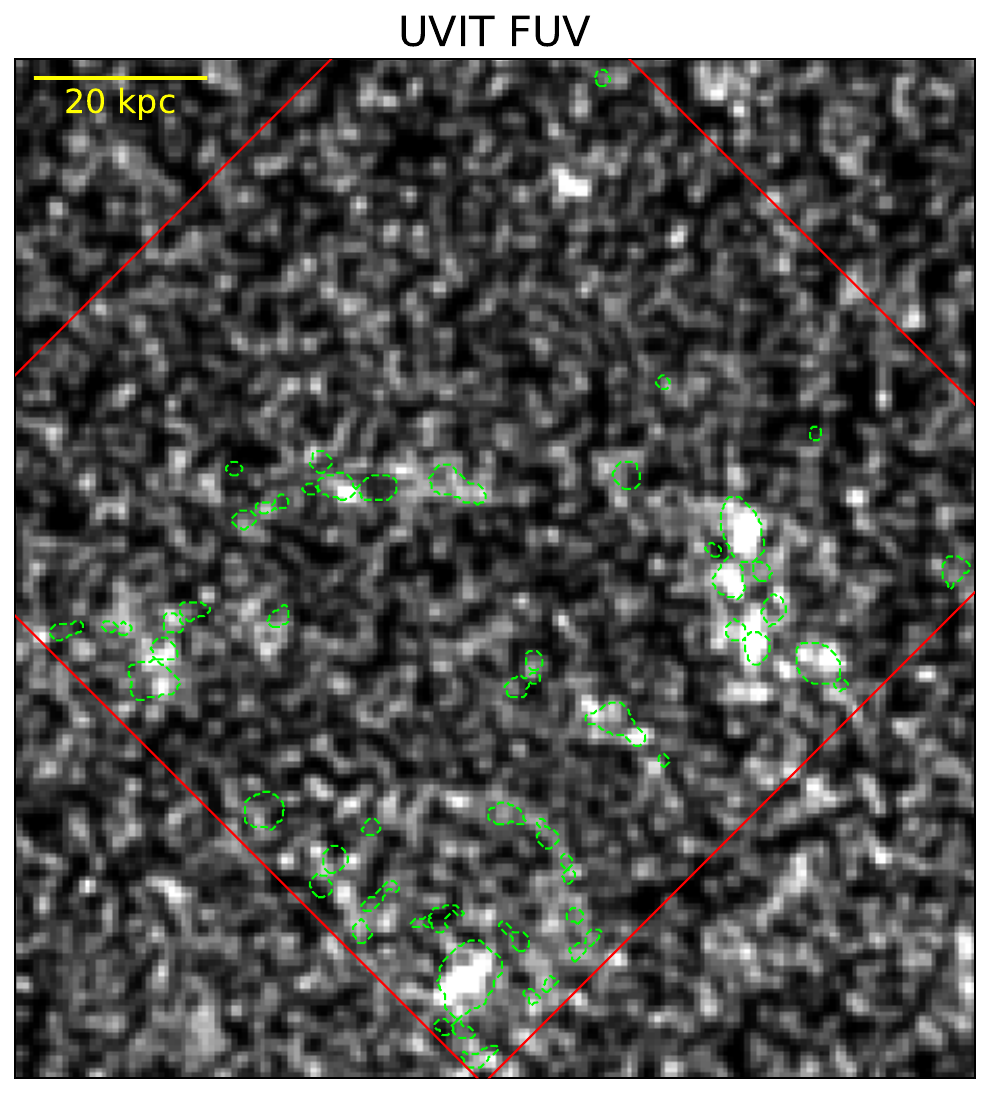}
    \caption{FUV image of Malin 1 from \citet{Saha2021} of the same field as our MUSE observations (shown as the red box). The green contours mark the \halpha{}-detected regions as discussed in Sect. \ref{sect:ppxf_fit_extended_disk}.}
    \label{fig:uvit_fuv}
\end{figure}

Figure \ref{fig:sigmasfr} shows the \sigmasfr{} as a function of galactocentric radius. In this plot, we also show the average \sigmasfr{} estimated from the \halpha{} radial average measurements discussed in Sect. \ref{sect:ppxf_halpha_radial_average}.
While the local measurements are indicative of individual star-forming regions, radial averages are meaningful to understand the galaxy evolution over orbital time-scales, to compare to 1-dimensional models depending on the galactocentric radius, or to the gas distribution over large scales.
We find a steep gradient in \sigmasfr{} within the central 10 kpc of the galaxy. This is consistent with the long-slit observations of Malin 1 from \citet{Junais2020}. Beyond the central regions, the \sigmasfr{} is mostly flat at about \sigmasfr{ $\sim 10^{-4}$} \msunyrkpcsq{} for the extended disk regions, but based on the radial average measurements there is a shallow decline to about $\sim 10^{-6}$ \msunyrkpcsq{} until 100 kpc. However, we see a clear spike in \sigmasfr{} around 50 to 60 kpc radius, in both the H$\alpha$ selected regions and the radial average estimates. This corresponds to the extended bright star-forming regions found at this radius as clearly seen in the \halpha{} map from Fig. \ref{fig:Gaussian_maps}. These individual star-forming regions have in general larger \sigmasfr{} than the radial average values. 
For comparison, we estimated a similar radial average of \sigmasfr{} using the UVIT FUV image of Malin 1 from \citet{Saha2021}, in the same field of our \halpha{} observations, and we obtain a similar peak.\footnote{We applied an attenuation correction for both the UV and the \halpha{} radial average data using a mean $A_{H\alpha}=0.36$ mag as obtained from Sect. \ref{sect:balmer_ratio}}. The attenuation at the FUV wavelength was obtained by adopting a \citet{calzetti2000} attenuation and a gas-to-stellar reddening factor of 0.44 as discussed in Sect. \ref{sect:discussion}. From Fig. \ref{fig:uvit_fuv} we can clearly see that most of the bright \halpha{} regions coincide well with the UV-blobs. These UV blobs were not resolved in the previous GALEX images of Malin 1 from \citet{boissier2016}. With the improved angular resolution of the UVIT images ($\sim$1.6\arcsec{}; \citealt{Saha2021}), which is three times better than that of GALEX, many \halpha{}-bright regions we observe can be identified as resolved individual regions in the UVIT image. Similarly, the \sigmasfr{} values based on the UV data are well consistent with the estimates from the \halpha{} measurements within their uncertainties (see Fig. \ref{fig:sigmasfr}), although many UV radial points only provide an upper limit in \sigmasfr{} while the \halpha{} determination is well constrained due to our spectral stacking technique discussed in Sect. \ref{sect:ppxf_halpha_radial_average}. We also performed a UV radial average measurement of \sigmasfr{} for the full galaxy (instead of only the one-quarter where we have MUSE observations) and found that they are very similar to our initial estimates with a difference of less than 0.1 dex. 

\subsection{Metallicity}\label{sect:metallicity}

We estimated the radial variation in metallicity in Malin 1 using the observed emission lines. We use the N2 and the O3N2 metallicity calibrators from \citet{Marino2013}, given as: 

\begin{equation}
12 + \log(O/H) = 8.743 + 0.462 \,N2    
\end{equation}

\begin{equation}
12 + \log(O/H) = 8.533 - 0.214 \,O3N2,    
\end{equation}

where the flux ratios are encoded as $N2 = \log([NII]_{6583}/H\alpha)$ and $O3N2 = \log([OIII]_{5007}/H\beta) - \log([NII]_{6583}/H\alpha)$. \\

\begin{figure*}
    \centering
    \includegraphics[width=0.49\textwidth]{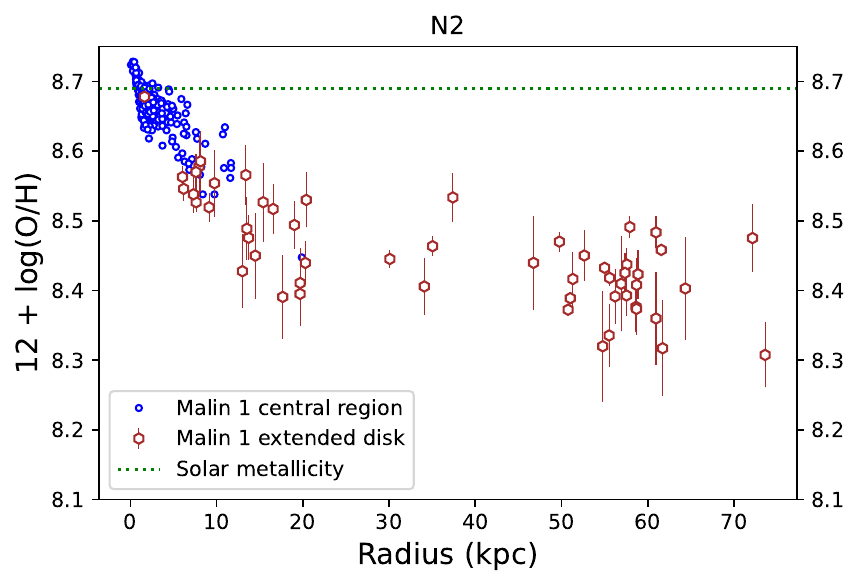}
    \includegraphics[width=0.49\textwidth]{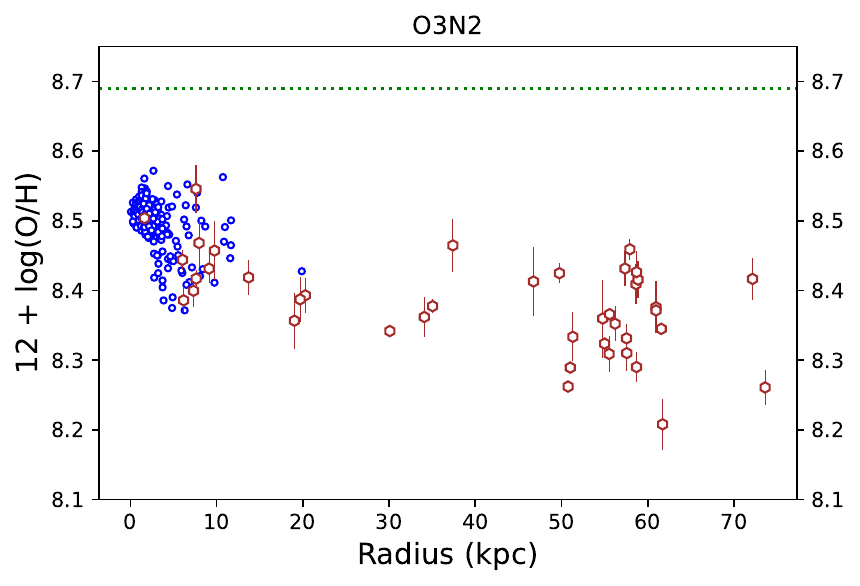}
    \caption{Radial variation of the metallicity of Malin 1 using the N2 calibrator (left panel) and the O3N2 calibrator (right panel), based on \citet{Marino2013}. The blue circles and the brown hexagons are the central regions and the extended disk regions, respectively, as discussed in Sect. \ref{sect:ppxf_fit}. The green horizontal dotted line marks the solar metallicity. Note that the N2 and O3N2 calibrations from \citet{Marino2013} have an additional calibration uncertainty of 0.16 dex and 0.18 dex, respectively. 
    }
    \label{fig:metallicity}
\end{figure*}

Figure \ref{fig:metallicity} shows the radial metallicity distribution of Malin~1. The metallicity estimate based on the N2 calibrator indicates that the central region of the galaxy (within a few kpc) has nearly solar metallicity, whereas, in the inner disk out to 20 kpc, we see a steep gradient in metallicity that reaches sub-solar values ($\sim$0.65 $Z_{\odot}$). However, for the outer disk beyond 20 kpc, we see a flattening in the metallicity ($\sim$0.6 $Z_{\odot}$),  consistent with no or a shallow slope, compared to the central regions. 
Such a behavior is also found in XUV disk galaxies like M83 \citep{Bresolin2009,Bresolin2017} and NGC\,1512 \citep{lopez-sanchez2015}. Both the N2 and the O3N2 calibrators show a similar trend, although in general, the metallicity estimated from O3N2 is lower than the N2 estimates by on average about 0.07 dex. Such an offset among different strong-line calibrations is often found in the literature as a result of differences in the excitation parameter and ionization states of the various lines used \citep[e.g.,][]{Kewley2008,Micheva2022}. For instance, \citet{Kewley2008} show that offsets among different calibrations can go up to 0.6 dex in metallicity, and with a large scatter. This is consistent with the offset of 0.07 dex we obtained between our N2 and O3N2 estimates. Other commonly used metallicity calibrators in the literature such as R$_{23}$ or N2O2 cannot be estimated using our data as the emission lines required for those calibrators are not in the MUSE spectral coverage. 

Table \ref{Table:region_properties} provides the measured quantities of all the \halpha{}-selected regions in the extended disk of Malin 1 discussed in this section. Based on the \halpha{} fluxes from Table \ref{Table:region_properties}, it is interesting to note that we observe \halpha{} luminosities ($L_{H\alpha}$) in the range of $10^{38}$ to $10^{40}$ erg s$^{-1}$ (with a median $L_{H\alpha}$ of $10^{38.7}$ erg s$^{-1}$), similar to the range found for HII regions of LSB galaxies by \citet{Schombert2013}. However, due to the limited resolution we have, where we cannot resolve individual HII regions at the distance of Malin 1, it is hard to make a direct comparison.

\begin{table*}[h]
\centering
\caption{Properties of the \halpha{} selected regions in the extended disk of Malin 1 as discussed in Sect. \ref{sect:ppxf_fit_extended_disk}.}\label{Table:region_properties}
\scalebox{0.9}{
\begin{tabular}{|*{9}{c|}}
\hline
Region & R.A & Dec. & Distance & Area & \halpha{} & \halpha{}/\hbeta & $12+\log O/H$ & $12+\log O/H$ \\
 & (deg) & (deg) & (kpc) & (arcsec$^{2}$) &  &  & [N2] & [O3N2]  \\
(1) & (2) & (3) & (4) & (5) & (6) & (7) & (8) & (9) \\
 \hline
1 & 189.2470 & 14.3287 & 9.13 & 1.80 & $1.11 \pm 0.04$ & $2.95 \pm 0.45$ & $8.52 \pm 0.02$ & $8.43 \pm 0.02$ \\
2 & 189.2474 & 14.3292 & 6.05 & 1.00 & $1.48 \pm 0.05$ & $2.60 \pm 0.28$ & $8.56 \pm 0.02$ & $8.44 \pm 0.01$ \\
3 & 189.2478 & 14.3293 & 6.18 & 0.84 & $3.29 \pm 0.11$ & $3.65 \pm 0.55$ & $8.55 \pm 0.02$ & $8.39 \pm 0.02$ \\
\textbf{4} & \textbf{189.2473} & \textbf{14.3303} & \textbf{1.64} & \textbf{17.04} & $\mathbf{26.07 \pm 0.15}$ & $\mathbf{3.23 \pm 0.06}$ & $\mathbf{8.68 \pm 0.00}$ & $\mathbf{8.50 \pm 0.00}$ \\
5 & 189.2459 & 14.3299 & 7.96 & 0.52 & $0.71 \pm 0.06$ & $2.02 \pm 0.45$ & $8.58 \pm 0.04$ & $8.47 \pm 0.03$ \\
6 & 189.2455 & 14.3302 & 9.76 & 0.52 & $0.61 \pm 0.05$ & $2.35 \pm 0.66$ & $8.55 \pm 0.05$ & $8.46 \pm 0.04$ \\
7 & 189.2449 & 14.3309 & 13.39 & 0.80 & $0.57 \pm 0.05$ & $2.52 \pm 0.59$ & $8.57 \pm 0.04$ & -- \\
8 & 189.2462 & 14.3311 & 7.63 & 1.12 & $1.69 \pm 0.04$ & $2.89 \pm 0.25$ & $8.53 \pm 0.01$ & $8.42 \pm 0.01$ \\
9 & 189.2446 & 14.3312 & 15.40 & 0.64 & $0.46 \pm 0.04$ & $1.94 \pm 0.46$ & $8.53 \pm 0.06$ & -- \\
10 & 189.2495 & 14.3313 & 12.99 & 1.28 & $0.57 \pm 0.03$ & -- & $8.43 \pm 0.05$ & -- \\
11 & 189.2465 & 14.3313 & 7.32 & 0.40 & $2.85 \pm 0.14$ & $3.69 \pm 0.69$ & $8.54 \pm 0.03$ & $8.40 \pm 0.02$ \\
12 & 189.2478 & 14.3316 & 7.61 & 1.92 & $0.82 \pm 0.04$ & $2.11 \pm 0.27$ & $8.57 \pm 0.03$ & $8.55 \pm 0.03$ \\
13 & 189.2483 & 14.3315 & 8.13 & 0.48 & $0.55 \pm 0.05$ & $1.83 \pm 0.42$ & $8.59 \pm 0.04$ & -- \\
14 & 189.2450 & 14.3316 & 14.48 & 0.76 & $0.81 \pm 0.06$ & $3.30 \pm 1.07$ & $8.45 \pm 0.06$ & -- \\
15 & 189.2493 & 14.3319 & 13.72 & 1.04 & $0.88 \pm 0.04$ & $2.76 \pm 0.41$ & $8.48 \pm 0.03$ & $8.42 \pm 0.02$ \\
16 & 189.2504 & 14.3322 & 19.69 & 1.64 & $0.53 \pm 0.03$ & -- & $8.41 \pm 0.05$ & -- \\
17 & 189.2489 & 14.3322 & 13.52 & 0.44 & $0.94 \pm 0.06$ & -- & $8.49 \pm 0.04$ & -- \\
18 & 189.2451 & 14.3324 & 16.56 & 0.48 & $0.94 \pm 0.06$ & -- & $8.52 \pm 0.04$ & -- \\
19 & 189.2501 & 14.3328 & 20.29 & 2.28 & $1.18 \pm 0.04$ & $3.43 \pm 0.55$ & $8.44 \pm 0.03$ & $8.39 \pm 0.02$ \\
20 & 189.2452 & 14.3327 & 17.67 & 0.48 & $0.89 \pm 0.05$ & -- & $8.39 \pm 0.06$ & -- \\
21 & 189.2456 & 14.3332 & 19.02 & 1.68 & $2.98 \pm 0.15$ & $5.69 \pm 2.04$ & $8.49 \pm 0.03$ & $8.36 \pm 0.04$ \\
22 & 189.2493 & 14.3334 & 20.40 & 0.92 & $0.50 \pm 0.03$ & $2.78 \pm 0.71$ & $8.53 \pm 0.04$ & -- \\
23 & 189.2516 & 14.3338 & 30.08 & 5.48 & $1.68 \pm 0.03$ & $3.24 \pm 0.20$ & $8.45 \pm 0.01$ & $8.34 \pm 0.01$ \\
24 & 189.2465 & 14.3337 & 19.68 & 2.64 & $0.78 \pm 0.03$ & $3.19 \pm 0.55$ & $8.40 \pm 0.05$ & $8.39 \pm 0.03$ \\
25 & 189.2431 & 14.3348 & 33.79 & 0.36 & $1.09 \pm 0.10$ & $3.55 \pm 1.39$ & -- & -- \\
26 & 189.2441 & 14.3356 & 35.05 & 6.12 & $0.76 \pm 0.01$ & $2.92 \pm 0.21$ & $8.46 \pm 0.01$ & $8.38 \pm 0.01$ \\
28 & 189.2540 & 14.3364 & 49.75 & 7.24 & $1.41 \pm 0.03$ & $3.62 \pm 0.33$ & $8.47 \pm 0.01$ & $8.42 \pm 0.01$ \\
29 & 189.2462 & 14.3363 & 34.08 & 1.68 & $1.00 \pm 0.04$ & $3.45 \pm 0.63$ & $8.41 \pm 0.04$ & $8.36 \pm 0.03$ \\
30 & 189.2394 & 14.3363 & 54.77 & 0.44 & $3.78 \pm 0.21$ & $4.86 \pm 1.43$ & $8.32 \pm 0.08$ & $8.36 \pm 0.06$ \\
31 & 189.2459 & 14.3365 & 35.27 & 0.40 & $0.62 \pm 0.05$ & $2.96 \pm 0.99$ & -- & -- \\
32 & 189.2399 & 14.3368 & 54.99 & 7.12 & $4.93 \pm 0.03$ & $3.38 \pm 0.07$ & $8.43 \pm 0.00$ & $8.32 \pm 0.00$ \\
33 & 189.2459 & 14.3368 & 37.40 & 1.08 & $0.76 \pm 0.05$ & $3.21 \pm 0.74$ & $8.53 \pm 0.04$ & $8.46 \pm 0.04$ \\
34 & 189.2411 & 14.3371 & 51.01 & 2.84 & $6.66 \pm 0.04$ & $3.26 \pm 0.07$ & $8.39 \pm 0.01$ & $8.29 \pm 0.00$ \\
35 & 189.2537 & 14.3371 & 51.31 & 1.96 & $3.84 \pm 0.16$ & $5.78 \pm 1.66$ & $8.42 \pm 0.04$ & $8.33 \pm 0.04$ \\
36 & 189.2558 & 14.3375 & 60.95 & 1.68 & $0.49 \pm 0.03$ & $2.26 \pm 0.35$ & $8.36 \pm 0.07$ & $8.38 \pm 0.04$ \\
37 & 189.2416 & 14.3374 & 50.76 & 1.44 & $7.35 \pm 0.07$ & $3.35 \pm 0.10$ & $8.37 \pm 0.01$ & $8.26 \pm 0.00$ \\
38 & 189.2546 & 14.3375 & 56.21 & 0.52 & $1.56 \pm 0.06$ & $3.07 \pm 0.47$ & $8.39 \pm 0.04$ & $8.35 \pm 0.03$ \\
39 & 189.2549 & 14.3375 & 57.57 & 0.48 & $14.82 \pm 0.47$ & $6.82 \pm 1.64$ & $8.44 \pm 0.02$ & $8.31 \pm 0.03$ \\
40 & 189.2535 & 14.3376 & 52.92 & 1.28 & $0.65 \pm 0.04$ & $3.17 \pm 0.79$ & -- & -- \\
41 & 189.2513 & 14.3377 & 46.75 & 1.32 & $0.57 \pm 0.04$ & $2.80 \pm 0.77$ & $8.44 \pm 0.07$ & $8.41 \pm 0.05$ \\
42 & 189.2408 & 14.3379 & 55.51 & 2.32 & $1.09 \pm 0.04$ & $3.46 \pm 0.48$ & $8.34 \pm 0.05$ & $8.31 \pm 0.03$ \\
43 & 189.2531 & 14.3379 & 52.63 & 1.76 & $3.42 \pm 0.16$ & $5.84 \pm 1.99$ & $8.45 \pm 0.04$ & -- \\
44 & 189.2417 & 14.3385 & 55.58 & 4.48 & $1.63 \pm 0.02$ & $3.03 \pm 0.15$ & $8.42 \pm 0.01$ & $8.37 \pm 0.01$ \\
45 & 189.2370 & 14.3387 & 73.60 & 2.56 & $1.46 \pm 0.04$ & $3.34 \pm 0.43$ & $8.31 \pm 0.05$ & $8.26 \pm 0.03$ \\
46 & 189.2410 & 14.3387 & 57.92 & 1.12 & $1.44 \pm 0.04$ & $2.87 \pm 0.24$ & $8.49 \pm 0.02$ & $8.46 \pm 0.02$ \\
47 & 189.2415 & 14.3395 & 61.56 & 10.20 & $3.63 \pm 0.02$ & $3.14 \pm 0.06$ & $8.46 \pm 0.00$ & $8.34 \pm 0.00$ \\
48 & 189.2421 & 14.3391 & 56.91 & 0.56 & $0.68 \pm 0.05$ & $2.15 \pm 0.44$ & $8.41 \pm 0.07$ & -- \\
50 & 189.2520 & 14.3397 & 58.63 & 1.44 & $1.66 \pm 0.05$ & $3.61 \pm 0.51$ & $8.38 \pm 0.04$ & $8.41 \pm 0.03$ \\
51 & 189.2516 & 14.3400 & 58.87 & 0.56 & $1.13 \pm 0.05$ & $2.76 \pm 0.38$ & $8.42 \pm 0.04$ & $8.42 \pm 0.03$ \\
52 & 189.2512 & 14.3401 & 58.70 & 0.64 & $1.10 \pm 0.05$ & $2.86 \pm 0.41$ & $8.41 \pm 0.04$ & $8.43 \pm 0.03$ \\
53 & 189.2475 & 14.3404 & 57.36 & 5.92 & $0.66 \pm 0.02$ & $2.99 \pm 0.36$ & $8.43 \pm 0.03$ & $8.43 \pm 0.02$ \\
54 & 189.2500 & 14.3404 & 58.69 & 3.44 & $1.11 \pm 0.04$ & $3.48 \pm 0.48$ & $8.37 \pm 0.04$ & $8.29 \pm 0.02$ \\
55 & 189.2492 & 14.3404 & 57.53 & 3.64 & $1.51 \pm 0.04$ & $3.99 \pm 0.59$ & $8.39 \pm 0.03$ & $8.33 \pm 0.02$ \\
56 & 189.2506 & 14.3404 & 58.98 & 0.48 & $0.57 \pm 0.05$ & $2.09 \pm 0.50$ & -- & -- \\
57 & 189.2439 & 14.3406 & 60.95 & 2.68 & $1.15 \pm 0.04$ & $3.47 \pm 0.49$ & $8.48 \pm 0.02$ & $8.37 \pm 0.02$ \\
58 & 189.2522 & 14.3408 & 64.35 & 0.64 & $0.54 \pm 0.04$ & -- & $8.40 \pm 0.07$ & -- \\
59 & 189.2504 & 14.3409 & 61.72 & 1.48 & $0.83 \pm 0.04$ & $3.41 \pm 0.68$ & $8.32 \pm 0.07$ & $8.21 \pm 0.04$ \\
60 & 189.2399 & 14.3415 & 74.44 & 0.48 & $0.45 \pm 0.05$ & -- & -- & -- \\
61 & 189.2431 & 14.3426 & 72.14 & 0.52 & $0.64 \pm 0.04$ & $1.72 \pm 0.27$ & $8.48 \pm 0.05$ & $8.42 \pm 0.03$ \\
62 & 189.2444 & 14.3488 & 104.76 & 0.72 & $0.52 \pm 0.05$ & -- & -- & -- \\
\hline
\end{tabular}
}
\tablefoot{(1) ID of the region as shown in Fig. \ref{fig:Gaussian_maps}. Region 4, marked as bold, corresponds to the central region of Malin 1. Note that the regions with  ID 27 and 49 are excluded from this work as they do not have an SNR $>2.5$ in any of the emission lines (see Sect. \ref{sect:ppxf_fit_extended_disk}); (2-3) Sky coordinates of the region; (4) Distance of the region from the center of Malin 1; (5) Area of the region; (6) \halpha{} surface brightness of the region in units of 10$^{-17}$ erg s$^{-1}$ cm$^{-2}$ arcsec$^{-2}$.
The \halpha{} flux given here is corrected for Galactic extinction, underlying stellar absorption, and dust attenuation as discussed in Sect. \ref{sect:results}; (7) Mean Balmer ratio of the region; (8-9) Mean oxygen abundance of the region estimated using the \citet{Marino2013} N2 and O3N2 calibrators, respectively.}
\end{table*}

\section{Discussion}\label{sect:discussion}

\subsection{Dust attenuation in Malin 1}

Low surface brightness galaxies are generally considered to be dust poor \citep{hinz2007,rahman2007,Liang2010}. However, these results are based
on either very small samples or shallow data. Recently \citet{Junais2023} performed a large statistical analysis of dust content in 1003 LSBs using deep data and found that although a fraction of LSBs is dust poor, a small fraction of them ($\sim$4\%) contains high dust attenuation. However, they observed that these LSBs with high attenuation also possess similarities with the giant LSBs in terms of their average stellar mass surface density and surface brightness. This may indicate a higher dust attenuation in GLSBs.

Our dust attenuation measurements of Malin 1 show that this is indeed the case. We observe a non-negligible amount of dust attenuation in the central region as well as the extended disk of Malin 1. %
\citet{Junais2023} calibrated the variation of attenuation as a function of the stellar mass surface density. Their equation 3 predicts that a Malin 1-like galaxy with an average stellar mass surface density of $10^{7.9}$ \msunkpcsq{} should have an attenuation $A_V$ of about 0.33 mag. The mean \halpha{} attenuation we obtained for Malin 1 corresponds to nearly 0.36 mag. This is consistent with the scaling relation predictions from \citet{Junais2023}.

The extended disk of Malin 1 is undetected in \textit{Spitzer} \citep{hinz2007}, \textit{Herschel} \citep{boissier2016}, and WISE\footnote{The WISE image of Malin 1 was inspected using the Legacy Survey Viewer (\url{https://www.legacysurvey.org/viewer})} imaging. For instance, \citet{hinz2007} found that Malin 1 has a $1\sigma$ flux upper-limit of 10 mJy in the \textit{Spitzer} MIPS 160 $\mu $m observations.
Our observed measurements of attenuation in the disk of Malin 1 challenge these non-detections. 
However, by construction, our relatively high attenuation\footnote{Assuming Case B recombination, i.e. optically thin gas.} is found in star-forming regions that cover only a small fraction of the extended disk (see Fig. \ref{fig:Gaussian_maps}). The relationship between the gas and the stellar attenuation has been largely discussed in the literature. \citet{Calzetti1997} proposed a factor of 0.44 between the gas and the stellar reddening as $E(B-V)_{\rm star} = 0.44E(B-V)_{\rm gas}$. \citet{Lin2020} investigated this relation for a large sample of galaxies with a wide range of physical properties and found that such a relation varies with several galaxy properties. The low covering fraction of star-forming regions in Malin 1 could be related to the low attenuation suggested by optical broadband studies. 
This calls for deeper observations of Malin 1 at MIR and FIR wavelengths. Moreover, exploring the geometric distribution of the dust and stars could also provide insights into the measured high attenuation and current non-detections at IR wavelengths \citep[e.g.,][]{Hamed2023}. 

An alternative explanation of this discrepancy might be that the conditions in the star-forming regions of Malin~1 are characterized by partial self-absorption of Balmer photons synonymous with mildly optically thick conditions. Such conditions would increase the expected Balmer ratio and consequently decrease the expected dust attenuation, making it more consistent with previous IR measurements.

\subsection{Correlation of radial gas metallicity and stellar profile}

\begin{figure}
    \centering
    \includegraphics[width=0.49\textwidth]{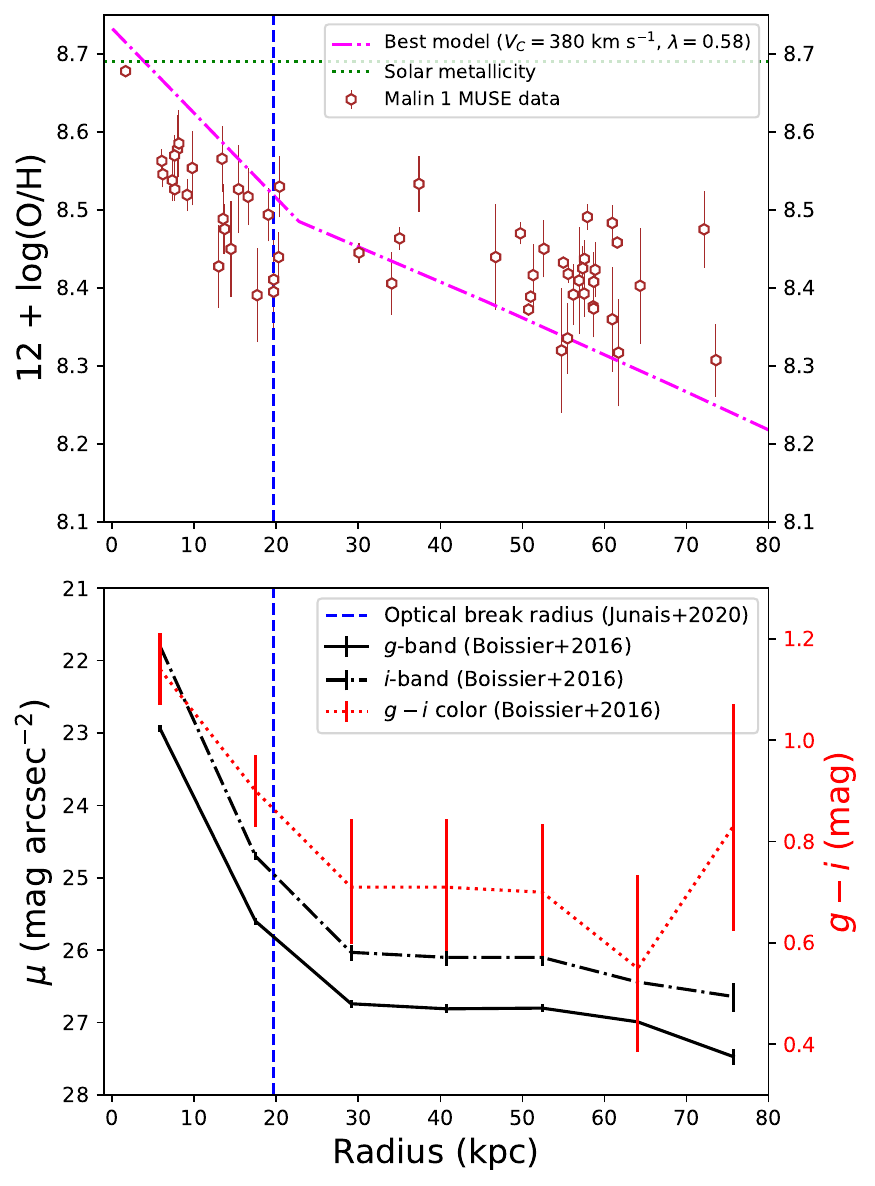}
    \caption{\textit{Top:} Metallicity gradient in Malin 1. 
    The brown hexagons are the Malin 1 extended disk regions discussed in Sect. \ref{sect:metallicity}. The observed metallicities shown here are based on the \citet{Marino2013} N2 calibrator. The magenta dot-dashed line is the best-fit model of Malin 1 with a circular velocity of 380 km s$^{-1}$ and a spin parameter of 0.58, obtained after a re-fitting following \citet{boissier2016}. \textit{Bottom}: Optical surface brightness profile of Malin 1 in the \textit{g}-band (black solid line) and the \textit{i}-band (black dot-dashed line) from \citet{boissier2016}. The secondary axis on the right shows $g-i$ color profile from \citet{boissier2016} as the red dotted line. Note that the stellar profiles are shown only until the radial range where we have a metallicity estimate. The vertical dashed blue line is the \textit{I}-band surface brightness break radius from \citet{Junais2020}}
    \label{fig:metallicity_comparison}
\end{figure}

Figure \ref{fig:metallicity_comparison} (top panel) shows the radial variation of gas-phase metallicity using the N2 calibrator in Malin 1. For the first time, we observe a relatively steep decline from the solar metallicity in the inner region of Malin 1 followed by a flattening of metallicity around 0.6 $Z_{\odot}$ beyond the 20 kpc radius. The radius at which this flattening occurs also coincides with the \textit{I}-band optical break radius of Malin 1 (19.6 kpc) found by \citet{Junais2020}. This latter break radius corresponds to the transition from the inner disk to the outer disk, based on a broken exponential disk profile \citep{Erwin2008}. This indicates that the inner and the outer disks of Malin 1 have different metallicity gradients. 

A flattening in metallicity beyond the break radius\footnote{The break radius of M83 is at 8.1 kpc \citep{Barnes2014}.} is also observed in galaxies like M83 \citep{Bresolin2009}. Interestingly, M83 is an XUV galaxy with a very extended UV disk. XUVs are thought to have similarities with GLSBs  \citep{thilker2007,bigiel2010,hagen2016}. Our observed similarities in the metallicity gradient support this hypothesis.
The flattening of the metallicity beyond the break radius could be due to several reasons. \citet{Bresolin2009} propose that such metallicity gradient could be the result of the flow of metals from the inner to the outer disk, accretion of pre-enriched gas, or due to past interaction with a satellite galaxy. 
We compared our observations in Malin 1 with the metallicity gradient from a  model similar to the one shown in \citet{boissier2016}. 
To this aim, we updated the \citet{boissier2016} model by performing a similar fit on the photometric profiles, but with the same grid of models as in \citet{Junais2022} (excluding ram-pressure). We find the best parameters to be very close to the original ones: $V_{C}= 380^{+200}_{-60}$ km s$^{-1}$ and $\lambda=0.58^{+0.18}_{-0.08}$, where $V_{C}$ and $\lambda$ are the circular velocity and the halo spin parameters, respectively from the models of \citet{boissier2016}. We do, however, find a difference as the metallicity is much higher than in the model published in \citet{boissier2016}. For instance, at a radius of 55 kpc in Malin 1, \citet{boissier2016} predicts a metallicity of 0.1 $Z_{\odot}$, whereas the best model we obtained now has a metallicity of 0.45 $Z_{\odot}$ at the same radius (see the dot-dashed magenta line in Fig. \ref{fig:metallicity_comparison}).
After inspection, we realized that the model in \citet{boissier2016} was actually using the \citet{Kroupa1993} initial mass function (IMF), while we adopted in \citet{Junais2022} the \citet{Kroupa2001} IMF. Differences in IMF can indeed generate large differences in the net yield integrated over the IMF \citep[e.g.,][]{Vincenzo2016}. The \citet{Kroupa1993} IMF being much poorer in massive stars than the one of \citet{Kroupa2001}, the metallicity profile here is about 5 times higher than the one published in \citet{boissier2016}. 
The overall metallicity level of the model is consistent with the observations within their scatter in the inner region at $\lesssim20$\,kpc (see Fig. \ref{fig:metallicity_comparison}), especially if we take into account the large systematic uncertainties of the model in the IMF and yields. However,
the metallicity gradient of the model fails to reproduce the metallicity plateau beyond 20\,kpc and is steeper than in the observations.
This indicates that this simple model (in which the system evolves in isolation without radial migration) is probably insufficient to reproduce all the properties of the galaxy (see e.g., \citealt{Kubryk2013} for the effect of radial migration on chemical abundance gradients). 

The bottom panel of Fig. \ref{fig:metallicity_comparison} shows the radial variation of the \textit{g}-band and \textit{i}-band optical surface brightness and color profiles of Malin 1 from \citet{boissier2016}, respectively. We can clearly notice a correlation in the radial surface brightness and color profiles with that of the metallicity gradient shown in the top panel of Fig. \ref{fig:metallicity_comparison}. For both the surface brightness and color profiles, we observe a steep decline in the inner part out to $\sim\!20$\,kpc, followed by a flattening, consistent with the trend in the metallicity. This is similar to the observation of \citet{Marino2016} for the extended disks of local spiral galaxies from the CALIFA survey. Based on the shape of the surface brightness profile, Malin 1 has an up-bending or anti-truncated Type III profile \citep{Erwin2005}, where the inner disk has a steeper surface brightness slope than the extended outer disk (see bottom panel of Fig. \ref{fig:metallicity_comparison}). \citet{Marino2016} found that Type III galaxies have a positive correlation between the change in color and metallicity gradient, followed by a flattening in metallicity. We find the same trend in Malin 1. There are several scenarios proposed in the literature on the formation of Type III profiles. \cite{Younger2007} found that minor mergers can produce Type III surface brightness profiles. On the other hand, \citet{Ruiz-Lara2017} suggested that Type III profiles form as a result of the radial migration of material from the inner to outer disk, as well as the accretion of material from the outskirts. This notion was challenged by \citet{Tang2020}, who found that stellar migration alone cannot form Type III profiles. Therefore, it is likely that most of the anti-truncated disk of Malin 1 resulted from the accretion of material from the outskirts (e.g., pre-enriched gas from a minor merger). 
The flattening of the color and metallicity profiles, and the rather high metallicity (0.6 $Z_{\odot}$) in the extended disk also point in this direction. Using the mass-metallicity scaling relations for gas and stars in isolated Local Group dwarf galaxies \citep[see][]{Hidalgo2017}, the outer disk gas abundance in Malin 1 would correspond to an LMC-type dwarf galaxy with a stellar and gas mass of $\sim\!10^{9.5-10}\,M_\odot$ each.

As gas metallicity traces recent star formation, photometric colors trace older stellar population ages (and abundances). The similarity of their radial profiles in Figure~\ref{fig:metallicity_comparison}, together with the break at 20 kpc in both, points to distinct evolutionary histories of the inner and outer galaxy systems over the epoch of the tracers, i.e. from Gyr in the past to the recent dozens of Myr traced by the gas. This in turn means that the mechanisms of star formation and feedback/enrichment in either of the components were not significantly influenced by the other, neither were both components altered by radial migration. In that sense, one could speak of Malin 1 as a composite galaxy with morphological sub-components that have evolved independently over many Gyr. The quantitative implications of this picture will be explored in future work.

The apparent correlation between the abundance and surface brightness profile also echoes the fact that abundance gradients expressed per disk scale length ($R_d$) tend to have a universal value with only a small scatter for disk galaxies \citep{Prantzos2000,Sanchez2014}. In Malin 1, we separated the inner disk ($<20$ kpc) and the outer disk ($>20$ kpc) as shown in Fig. \ref{fig:metallicity_comparison}. \citet{Junais2020} showed that the inner and outer disks of Malin 1 are separated at $\sim$ 20 kpc, with scale lengths of 5.3 kpc and 41.8 kpc, respectively. We performed separate linear fits to the abundance gradients of both disks. We found that, for the inner and the outer disk, the slope of the abundance gradient normalized by its disk scale length has a value of $-0.09$ dex/$R_d$ and $0.04$ dex/$R_d$, respectively. \citet{Sanchez2014} showed that for local disk galaxies, this value is in the range of $-0.06\pm0.05$ dex/$R_d$. Therefore, the normalized abundance gradient of the inner disk of Malin 1 has a negative value, close to the mean value found in normal galaxies. However, the outer disk is consistent with a small positive gradient, but within the $2\sigma$ scatter of \citet{Sanchez2014}, suggesting that the extended disk of Malin 1 lies along the extreme tail of ``standard'' galaxies. This may point toward a particular mode of star formation at relatively low densities.

\subsection{Star formation efficiency in the low-density regime}

\begin{figure}
    \centering
    \includegraphics[width=0.49\textwidth]{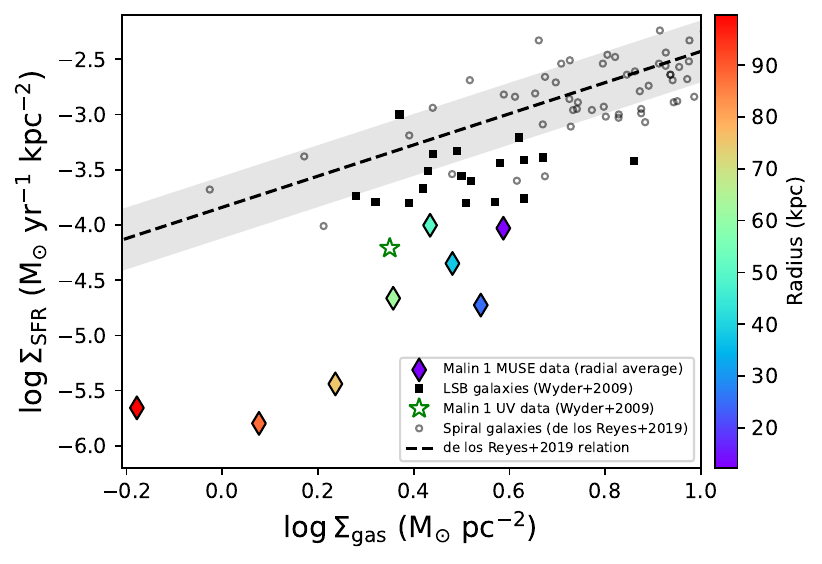}
    \caption{Star formation rate surface density versus gas surface density. The diamond symbols mark the region along the disk of Malin 1 based on the radial averaged \sigmasfr{} we estimated as in Fig. \ref{fig:sigmasfr}. The $\Sigma_{\rm gas}$ for Malin 1 is from the eight \Hi{} data points of \citet{Lelli2010}, after correcting for the Helium by a factor of 1.4. The color bar indicates the radius along the disk of Malin 1. The black open circles are normal spiral galaxies from \citealt{delosReyes2019} (the $\Sigma_{\rm gas}$ from \citealt{delosReyes2019} uses the total atomic and molecular gas mass). The black dashed line and the grey shaded region are the \citet{delosReyes2019} best-fit relation and $1\sigma$ scatter. The black squares are the LSB galaxies from \citet{Wyder2009}, along the Malin 1 marked as the open green star symbol based on the UV \sigmasfr{} estimate from \citet{Wyder2009}. 
    }
    \label{fig:KS_plot}
\end{figure}

The efficiency of star formation activity in the low-density regime is often debated. %
LSBs are believed to have very low star formation efficiencies compared to their HSB counterparts \citep[e.g.,][]{Bigiel2008,Wyder2009}. This results in the divergence of LSBs from the global Kennicutt-Schmidt law (KS relation; \citealt{Kennicutt1998}) for star-forming galaxies as shown in Fig. \ref{fig:KS_plot}. This may be related to the possible existence of a gas density threshold \citep{delosReyes2019}, below which  
star formation has a very low efficiency. %
The extended disks of GLSB galaxies offer a new laboratory to study the star formation activity in this low-density regime.

Figure \ref{fig:KS_plot} shows our measured \sigmasfr{} and gas surface density for Malin 1 at various radii of its disk. For this, we used the \Hi{} gas profile from \citet{Lelli2010}, which has been obtained from VLA data, with a resolution of $\sim$20\arcsec. We combined these profiles with our azimuthally averaged SFR profile from Sect. \ref{sect:sigma_sfr}. Even if our data is only obtained in one-quarter of the galaxy, we can guess this comparison because the gas distribution is expected to be symmetric (see Fig. 6 of \citealt{Lelli2010}).
We see that, at a given \Hi{} gas surface density, $\Sigma_{\rm gas}$, the \sigmasfr{} level of Malin 1 falls much below the level and scatter expected for normal star-forming spiral galaxies from \citet{delosReyes2019}. This indicates that the Malin 1 disk has a very low star formation efficiency. This effect is dominant in the outermost part of the disk ($r>60$ kpc) where we see a large difference of $\sim$2 dex in \sigmasfr{} with respect to normal star-forming galaxies, whereas it is $\sim$1 dex for the inner regions. Within this region, the \sigmasfr{} of Malin 1 is also consistent with what is observed for other LSBs and GLSBs in the literature \citep{Boissier2008,Wyder2009,Saburova2021}. Our estimates agree with the average value of \sigmasfr{} for Malin 1 found by \citet{Wyder2009} based on the GALEX UV data. Overall, we can clearly see that the Malin 1 disk lies along the extreme end of \sigmasfr{} level than for any other known LSB galaxy.

We point out that the gas surface density of normal spiral galaxies shown in Fig. \ref{fig:KS_plot} from \citet{delosReyes2019} includes the total of atomic \Hi{} and molecular H$_{2}$ gas whereas the LSBs from \citet{Wyder2009} and our Malin 1 data points only include \Hi{}. However, it is reasonable to assume that the molecular gas fraction is negligible in LSBs and GLSBs \citep{Braine2000, galaz2008, Wyder2009}. Moreover, any additional H$_{2}$ gas in Malin 1 would move the points shown in Fig. \ref{fig:KS_plot} towards the right, i.e. further away from the relation for normal galaxies. The fact that, on the one hand, we are detecting huge regions of young stars surrounded by ionized gas emitting in H$\alpha$, and on the other hand, no CO emission was detected in the past efforts, neither recent ones with mm observations, points toward a quite peculiar interstellar medium (ISM) in Malin 1. As suggested by some authors \citep[][and references therein]{Galaz2022}, not only could the ISM be at a very low density, as all these studies seem to indicate, but also at a higher temperature than the ISM observed in high surface brightness spirals.
Following \citet{Boissier2004}, we made an estimate of the gas-to-dust ratio in the extended disk of Malin 1 by using the ratio of our observed \textit{V}-band attenuation A$_{\rm V}$ (from Eq. \ref{eq:balmer_attenuation}) and the \Hi{} column density ($N_{H}$) from \citet{Lelli2010}. We found a mean gas-to-dust ratio of $\log(M_{\rm gas}/M_{\rm dust}) = 2.06$. For our average gas metallicity of $\sim$0.6 $Z_{\odot}$ in the extended disk, such a gas-to-dust ratio is consistent with the lower limit, but within the scatter found for normal galaxies \citep[e.g.,][]{Boissier2004,remy-ruyer2014}.

\section{Conclusions}\label{sect:conclusions}

In this work, we present VLT/MUSE IFU observations of the giant low surface brightness galaxy Malin 1. We extracted several ionized gas emission lines using this data and performed a detailed analysis of the star formation rate, dust attenuation, and gas metallicity within this galaxy.

Our main results are summarized as follows.

\begin{itemize}

    \item For the first time, we observe strong \halpha{} emission in numerous regions along the extended disk of Malin 1 up to a radial distance of $\sim$100 kpc. Other emission lines (\nii{}, \hbeta{}  and \oiii{}) were also observed, but in fewer regions. This indicates that recent star formation is ongoing in several regions of the large diffuse disk of Malin 1.
    
     \item We estimated the Balmer decrement and dust attenuation in several regions of the galaxy and found that the Malin 1 disk has a mean Balmer ratio of 3.26 and an \halpha{} attenuation of 0.36 mag, assuming case-B gas conditions. This is also true at several tens of kpc from the galaxy center where we measured it in a bright star-forming region.

    \item Malin 1 has a steep decline in the star formation rate surface density (\sigmasfr{}) within the inner 20 kpc, followed by shallow decline in the extended disk. We also see a peak in \sigmasfr{} around the 60 kpc radius. Our radial average \sigmasfr{} estimates based on \halpha{} are consistent with the measurements from UV as well as other works from the literature. 

    \item The gas metallicity in Malin 1 shows a steep decline in the central region, similar to the radial \halpha{} profile. However, we observe a flattening of the metallicity in the extended disk with a rather high value of around 0.6 $Z_{\odot}$ within a radius range between 20 kpc and 80 kpc. We found that the outer disk abundance gradient in Malin 1, normalized by its scale length has a value close to zero, which is flatter than most normal disk galaxies in the literature. The abundance measurements in these very extended regions confirm that the gas is not primordial and that the gradient is flatter than expected for very simple models. Together with similar radial trends from photometric colors, this result suggests distinct star formation histories for the inner and outer disks of Malin 1 with little radial migration or interactions at play during the formation of its very extended disk.
    \item Comparison of our estimated star formation rate surface density and the gas surface density shows that, unlike normal spiral galaxies, Malin 1 lies along the regime of very low star formation efficiency, as found in other LSBs, but at the extreme lower limit. 
 
\end{itemize}

\begin{acknowledgements}
J and KM are grateful for support from the Polish National Science Centre via grant UMO-2018/30/E/ST9/00082. PMW gratefully acknowledges support by the German BMBF from the ErUM program (project VLT-BlueMUSE, grant 05A20BAB). GG, EJJ, and THP gratefully acknowledge support by the ANID BASAL project FB210003. THP gratefully acknowledges support through a FONDECYT Regular grant (No. 1201016). EJJ acknowledges support from FONDECYT Iniciaci\'on en investigaci\'on 2020 Project 11200263.
\end{acknowledgements}

\bibliographystyle{aa}
\bibliography{bibliography}

\end{document}